\documentclass[a4paper,12pt,autodetect-engine]{article}

\usepackage{amsmath}
\usepackage{amssymb}
\usepackage{amsfonts}
\usepackage{amsthm}
\usepackage{graphicx,hyperref,color} 
\usepackage{cite}
\usepackage{cancel} 

\newcommand{\del}{\partial}

 \makeatletter
 \@addtoreset{equation}{section}
 \makeatother
\usepackage{ytableau} 
\ytableausetup{smalltableaux}
\ytableausetup{centertableaux}
\newcommand{\dop}{\mathrm{d}}
\newcommand{\LeftD}{{\dop}}
\newcommand{\RightD}{\tilde{{\dop}}}
\newcommand{\group}[1]{\mathrm{#1}}
\newcommand{\GaussLaw}{Gauss's law}

\date{empty}
\begin{document}
\begin{titlepage}
\null
\begin{flushright}
May, 2025
\end{flushright}
\vskip 2cm
\begin{center}
{\large \bf 
Electric-Magnetic Duality for Symmetric Tensor Gauge Theories
\\
\vspace{0.2cm}
and Immobile $p$-branes
}
\vskip 2cm
\normalsize
\renewcommand\thefootnote{\alph{footnote}}

{\large
Ryuki Makino${}^{1}$\footnote{makino.ryuki(at)st.kitasato-u.ac.jp}
,
Shin Sasaki${}^{1}$\footnote{shin-s(at)kitasato-u.ac.jp}
and
Kenta Shiozawa${}^{2}$\footnote{shiozawa.kenta(at)kitasato-u.ac.jp}
}
\vskip 0.5cm

  {\it
  ${}^{1}$Department of Physics, Kitasato University \\
  Sagamihara 252-0373, Japan
  }

\vskip .1cm

  {\it
  ${}^{2}$Center for Natural Sciences, College of Liberal Arts and
  Sciences, \\ 
  Kitasato University \\
  Sagamihara 252-0373, Japan
  }
\vskip 0.7cm
\begin{abstract}
We study electric-magnetic duality in Lorentz invariant symmetric
 tensor gauge theories, where immobile charged
 particles---fractons---arise due to 
the generalized current conservation $\del_{\mu} \del_{\nu} J^{\mu \nu}
 = 0$
 and the fracton gauge principle.
We show that the duality in the symmetric gauge theories holds only in
 four-dimensional spacetime.
In higher dimensions, the duality does not hold with only the symmetric
 gauge fields but tensor fields with more complex symmetries come into
 play.
Furthermore, we show that a hierarchy for the symmetric gauge field theories
 of higher ranks is interpreted by the bi-form calculus.
We also discuss the restricted immobility of $p$-branes in the mixed symmetric gauge theories.
As a byproduct, we find that novel self-duality conditions are defined
 as BPS equations in the four-dimensional Euclidean space.

\end{abstract}
\end{center}

\end{titlepage}

\newpage
\setcounter{footnote}{0}
\renewcommand\thefootnote{*\arabic{footnote}}
\pagenumbering{arabic}
\tableofcontents

\section{Introduction} \label{sect:introduction}

Recent studies on fractons \textit{i.e.}\ immobile charged particles together
with the conserved dipole moments, in condensed matter physics 
\cite{Nandkishore:2018sel, Pretko:2020cko,
Vijay:2016phm,Bulmash:2018knk, Gromov:2022cxa, Radzihovsky:2019afv} 
had a profound impact on high-energy physics.
The fracton phase is characterized by the charged excitations of restricted mobility.
Namely, particles can freely move in a subspace of spacetime but they
are forbidden to escape from this space.
This is a consequence of the conserved dipole moment.
The fractons have been first discussed in the lattice field theory
\cite{Vijay:2016phm,Haah:2011drr}.
It has been shown that continuum limits of the lattice theory are given
by non-relativistic BF theories \cite{Slagle:2018swq, Slagle:2020ugk, Seiberg:2020cxy}.
Dualities of the non-relativistic field theories have been also studied
\cite{Seiberg:2020bhn, Seiberg:2020wsg, Gorantla:2020xap}.

Based on the fractonic nature of charged particles, the so-called
fracton gauge principle has been proposed \cite{Pretko:2018jbi}.
In this theory, the symmetric gauge field $A_{ij} = A_{ji}$ that is
subject to the gauge transformation of the form $\delta A_{ij} = \del_i \del_j
\lambda$ where $\lambda = \lambda (x)$ is the gauge parameter, plays a
central role.
The non-relativistic gauge theory that governs the fracton dynamics have been developed
\cite{Pretko:2016kxt, Bulmash:2018lid, Hirono:2022dci}.

Recently, a Lorentz invariant theory of fractons based on the
rank-2 symmetric tensor gauge field $A_{\mu \nu} = A_{\nu \mu}$ has been established
 \cite{Bertolini:2022ijb}.
The immobility of charged particles is implemented by the conservation of
a higher rank current $J^{\mu \nu}$. 
This is also expressed by a {\GaussLaw} $\del_i \del_j E^{ij} = \rho$ derived
from the equations of motion for the gauge field $A_{\mu \nu}$.
Since gauge theories based on the fracton gauge principle differ from
conventional gauge theories in many aspects, it is important to
investigate the fundamental properties of the symmetric gauge theories.
Several topics on the Lorentz invariant theory including 
the self-duality of the symmetric Maxwell theory \cite{Bertolini:2025jov},
the anomaly of the symmetric gauge field theory \cite{Rovere:2024nwc},
the relation to the linearized gravity \cite{Pretko:2017fbf},
the extension to the higher rank symmetric gauge theories and multipoles \cite{Wang:2019cbj},
the $(n,k)$ fractonic gauge theory \cite{Shenoy:2019},
and applications to extended objects \cite{Bertolini:2024jen}
have been investigated.

The purpose of this paper is to investigate an electric-magnetic duality
for symmetric gauge field theories.
It is well-known that the $p$-form electrodynamics in $D$ spacetime
dimensions are dual to the $(D-p-2)$-form electrodynamics.
The equation of motion and the Bianchi identity are interchanged in the
process of the duality.
This is easily understood by the properties of the Hodge duality for differential forms.
However, unlike the conventional $p$-form electrodynamics, the symmetric gauge theory
is not based on the differential forms in the usual sense.
As a result, the ``Bianchi identity'' of the gauge field is not
expressed geometrically and the systematic treatment of duality
transformation by the Hodge star operator can not be applied.
Therefore it is highly non-trivial to find the dual theories for a
given symmetric gauge theory of arbitrary ranks.
We explicitly construct the electric-magnetic duality between different
symmetric gauge theories by performing direct calculations based on the
``Bianchi identity'' derived from the fracton gauge principle. 
Furthermore, we show that this construction indeed realizes an
electric-magnetic duality within the framework of the higher rank symmetric gauge
theories.
In this process, depending on the rank of the symmetric gauge field and
spacetime dimensions, we will find dualities between the symmetric gauge theories and 
theories of mixed symmetric tensor gauge fields especially in higher
dimensions.
The situation is quite similar to the dual graviton theories in higher
dimensions where mixed symmetric tensor gauge fields naturally appear
\cite{Hull:2000zn, Hull:2001iu, Casini:2003kf, Hull:2024qpy,
Hull:2024bcl}.

The organization of this paper is as follows.
In the next section, we briefly introduce the symmetric gauge theory of rank-2 
based on the fracton gauge principle \cite{Pretko:2018jbi, Bertolini:2022ijb} and discuss
the immobility of charged particles.
We will comment on the higher rank generalizations of the symmetric gauge
theory and the multipole immobility.
We will also discuss a hierarchical structure of the field strength for 
the symmetric gauge fields and the bi-form calculus
\cite{Francia:2004lbf, deMedeiros:2002qpr, Hinterbichler:2022agn}.
In Section \ref{sec:EM_dual}, based on the Bianchi identities in the
symmetric gauge theories, we study the electric-magnetic
duality for the symmetric gauge theories in four-dimensional spacetime.
We show that the duality between symmetric gauge theories is possible
only in four dimensions.
In Section \ref{sec:duality_higher_dimensions}, we study the electric-magnetic
duality in diverse dimensions.
We find that the mixed symmetric (partially symmetric and partially
anti-symmetric) tensor gauge fields naturally appear in dimensions $D \ge 5$.
In lower dimensions $D \le 3$, we show that there are no non-trivial dualities.
In Section \ref{sec:immobile_p-branes}, 
we show that the mixed symmetric tensor gauge fields subject to the fracton
gauge principle couple to the 
mixed symmetric tensor currents
associated with charges of
the extended objects, \textit{i.e.}\ $p$-branes.
The dualities in gauge theories imply the fractonic particle/$p$-brane dualities in
higher dimensions.
In Section \ref{sec:BPS}, we discuss the self-duality conditions for the
symmetric gauge theories in four dimensions. 
We find that this is obtained as the BPS equation in Euclidean space.
Section \ref{sec:conclusion} is devoted to the conclusion and discussions.
In Appendix \ref{sec:dual_vector}, we discuss the dual of the vector type gauge theory.

\section{Symmetric gauge theory and fracton}
In this section, we briefly introduce the symmetric gauge theory that
implements conserved dipole moments.
This naturally leads to immobile charged particles called fractons.
We then generalize the formalism to higher rank symmetric gauge theories
that implement conservations of the dipole moments and immobilities of multipoles.

\subsection{Symmetric gauge theory}
The symmetric tensor gauge theory of rank-2 in four-dimensional spacetime is defined by the
following Lagrangian\footnote{
We use the mostly plus convention of the Minkowski metric $\eta_{\mu
\nu} = \text{diag} (-1, 1, \ldots, 1)$.
The indices $\mu, \nu, \ldots$ run over the whole spacetime while $i,j,
\ldots$ cover the spatial directions.
} \cite{Bertolini:2022ijb},
\begin{align}
\mathcal{L} = - \frac{1}{6} F_{\mu \nu \rho} F^{\mu \nu \rho},
\label{eq:symmetric_maxwell_Lagrangian}
\end{align}
where the field strength $F_{\mu \nu \rho}$ for the symmetric gauge
field $A_{\mu \nu} = A_{\nu \mu}$ is defined by
\begin{align}
F_{\mu \nu \rho} = \del_{\mu} A_{\nu \rho} + \del_{\nu} A_{\mu \rho} - 2
 \del_{\rho} A_{\mu \nu}.
\label{eq:field_strength}
\end{align}
Note that $F_{\mu \nu \rho} = F_{\nu \mu \rho}$ does not show any
symmetry in the last two indices but satisfies the cyclic identity and a ``Bianchi identity'':
\begin{align}
F_{\mu \nu \rho} + F_{\nu \rho \mu} + F_{\rho \mu \nu} = 0, \qquad 
\varepsilon^{\alpha\mu\nu\rho} \partial_\mu F_{\beta\nu\rho} = 0,
\label{eq:Bianchi_identity_for_rank-2_4d}
\end{align}
where $\varepsilon_{\mu \nu \rho \sigma}$ is the totally antisymmetric symbol. 
The Lagrangian \eqref{eq:symmetric_maxwell_Lagrangian} is invariant
under the following gauge transformation 
\begin{align}
\delta A_{\mu \nu} = \del_{\mu} \del_{\nu} \lambda,
\label{eq:scalar_gauge_transformation}
\end{align}
where $\lambda = \lambda (x)$ is the gauge parameter.
The gauge transformation \eqref{eq:scalar_gauge_transformation}
involving an extra derivative is characteristic of the fracton gauge
principle \cite{Pretko:2018jbi}.
The gauge field $A_{\mu \nu}$ naturally couples to the current $J^{\mu
\nu} = J^{\nu \mu}$ via $\mathcal{L}_{J} = - A_{\mu \nu} J^{\mu \nu}$.
By including the source term, the equation of motion for $A_{\mu \nu}$ is
given by
\begin{align}
\del_{\rho} F^{\mu \nu \rho} = -J^{\mu \nu}.
\end{align}
This is a generalization of the Maxwell equation of the ordinary
electromagnetic theory.

Now we define the electric and the magnetic fields as follows:
\begin{align}
    E^{ij}=F^{ij0},
    \qquad
    B_i^{\:\:j}=\frac{2}{3}\varepsilon_{0ikl}F^{jkl}.
\end{align}
Due to the cyclic identity in \eqref{eq:Bianchi_identity_for_rank-2_4d},
we have the relation
\begin{align}
    \del_i \del_j F^{ij0} = - 2 \del_i \del_j F^{0ij}.
\end{align}
Then, from the equation of motion, we have the {\GaussLaw}:
\begin{align}
    \del_i \del_j E^{ij} = \rho.
\end{align}
Here the charge density $\rho$ is defined by
\begin{align}
2\del_iJ^{0i}=\rho.
\end{align}
We also have the current conservation from the equation of motion:
\begin{align}
    \del_{\mu}\del_{\nu}J^{\mu\nu}=0.
\end{align}
This implies the relation
\begin{align}
    \del_0\rho+\del_i\del_jJ^{ij}=0,
\end{align}
where we have assumed $\partial_0\partial_0J^{00} = 0$.\footnote{
This assumption is naturally understood by the following reason.
The current $J^{\mu\nu}$ is typically expressed in terms of a complex
scalar field $\Phi$ as
$J^{\mu\nu}=i\{\Phi^2(D^{\mu\nu}[\Phi^2])^{\dagger}-(\Phi^{\dagger})^2D^{\mu\nu}[\Phi^2]\},$
where the covariant derivative
$D_{\mu\nu}[\Phi^2] = \Phi
\del_{\mu}\del_{\nu}\Phi - \del_{\mu}\Phi\del_{\nu}\Phi-
iA_{\mu\nu}\Phi^2$
is defined by gauging the transformation $\Phi \mapsto e^{i x_{\mu} \lambda^{\mu}}
\Phi$ \cite{Pretko:2018jbi}. 
The invariance under the
transformation $x_{i}\lambda^{i}$ results in the conservation of the
spatial dipole moment, while invariance under $x_0\lambda^0$ implies the
conservation of the temporal dipole moment. 
Since the physical meaning of the temporal dipole moment is ambiguous, 
we assume that $\Phi$ is independent of $x^0$ and $\del_0 \del_0 A^{00} = 0$.
}
This continuity relation results in the following conservation laws:
\begin{align}
    \begin{split}
    \del_0 Q & = \del_0 \int\! {\dop}V \,\rho=-\int\! {\dop}V\,\del_i\del_jJ^{ij}=0, \\
    \del_0 Q^k & =\del_0 \int\! {\dop}V \,x^k\rho=-\int\! {\dop}V \,x^k\del_i\del_jJ^{ij}=0,
    \end{split}
\end{align}
where $Q,Q^k$ are the total charge and the dipole moment in the spatial
volume $V$, respectively.
Therefore the symmetric tensor gauge field $A_{\mu\nu}$ naturally
introduces the conservations of $Q$ and $Q^k$.
These conservation laws restrict the mobility of the isolated charges
since the conservation of $Q^k$ forbids changes of the dipole moment.
However, this does not forbid the motion of a pair of charges (a dipole). 
They can move freely.
A particle with restricted mobility is called a fracton.

When we impose the traceless condition on the electric field $E^{ij}$, 
\begin{align}
        E^{i}{}_{i}=0,
\end{align}
then we have the condition,
\begin{align}
        Q^i{}_{i}=\int\! {\dop}V \,x^ix_i\,\rho=\int\! {\dop}V \,x^ix_i\del_k\del_l
 E^{kl}=-\int\! {\dop}V \,E^i{}_i=0.
\end{align}
Here $Q^{ij} = \int \! {\dop}V \, x^i x^j \rho$ is the total quadrupole moment.
This additional conservation law restricts the mobility of the dipole in
addition to the isolated charged particles.
In the material, a dipole can move by creating and annihilating quadrupoles in its vicinity.
In particular, a linear quadrupole makes the dipole move in a direction
parallel to the moment, while a square quadrupole makes the dipole move
in a direction perpendicular to the moment.
However, once the traceless condition is imposed on the electric field
$E^{ij}$, the trace of the quadrupole moment is preserved, and no linear
quadrupole is produced \cite{Pretko:2016kxt}.
This results in the restriction of the dipole motion to a two-dimensional space.
A particle with its motion restricted to two dimensions is called a planon.

Now we consider energy density associated with the action
\eqref{eq:symmetric_maxwell_Lagrangian}.
The stress tensor $T_{\alpha\beta}$ is defined by
\begin{align}
    T_{\alpha\beta}&=\left. - \frac{2}{\sqrt{-g}}\frac{\delta }{\delta
 g^{\alpha\beta}}\int\! {\dop}^4x\,\sqrt{-g}\left(-
 \frac{1}{6}F_{\mu\nu\rho}F^{\mu\nu\rho}\right)\right|_{g^{\alpha\beta}=\eta^{\alpha\beta}}.
\end{align}
The $T_{00}$ component is therefore given by
\begin{align}
T_{00}&=\frac{1}{4}(E^{ij}E_{ij}+B_i{}^jB^i{}_j).
\end{align}
This is nothing but a generalization of the energy density in Maxwell theory.

\subsection{Higher rank generalizations}
The aforementioned
relations among the fractons, the dipole moment and the
symmetric gauge field are easily generalized to the higher rank gauge theories.
We consider the symmetric gauge field of rank-$k$, $A_{\mu_1 \cdots \mu_k}$, and the
gauge transformation:
\begin{align}
\delta A_{\mu_1 \cdots \mu_k} = \del_{\mu_1} \cdots \del_{\mu_k} \lambda.
\end{align}
The gauge invariant field strength satisfying the cyclic identity 
$F_{\mu_1 \cdots \mu_{k+1}} + \text{c.p.} = 0$, where c.p.\ stands for the cyclic permutations, 
is defined by
\begin{align}
F_{\mu_1 \cdots \mu_{k+1}} = 
\del_{\mu_1} A_{\mu_2 \cdots \mu_{k+1}}
+
\del_{\mu_2} A_{\mu_1 \mu_3 \cdots \mu_{k+1}}
+
\cdots
+
\del_{\mu_{k}} A_{\mu_1 \cdots \mu_{k-1} \mu_{k+1}}
- 
k \,  
\del_{\mu_{k+1}} A_{\mu_1 \cdots \mu_k}.
\label{eq:rank-r_field_strength}
\end{align}
The gauge invariant Lagrangian is given by
\begin{align}
\mathcal{L} = - \frac{1}{k (k+1)} F_{\mu_1 \cdots \mu_{k + 1}} F^{\mu_1 \cdots \mu_{k+1}}.
\label{eq:Lagrangian_rank-k}
\end{align}
The symmetric gauge field $A_{\mu_1 \cdots \mu_k}$ naturally couples to the
symmetric current $J^{\mu_1 \cdots \mu_k}$.
This leads to the equation of motion:
\begin{align}
\del_{\nu} F^{\mu_1 \cdots \mu_k \nu} = -J^{\mu_1 \cdots \mu_k}.
\end{align}
Then the current conservation law $\del_{\mu_1} \cdots \del_{\mu_k} J^{\mu_1
\cdots \mu_k} = 0$ is easily follows.

The electric and the magnetic fields are defined as
\begin{align}
    E^{i_1\cdots i_k}=F^{i_1\cdots i_k 0},
    \qquad
    B_i^{\:\:j_1\cdots j_{k-1}}=\frac{k}{k+1} \varepsilon_{0iqp} F^{j_1\cdots j_{k-1}qp}.
\end{align}
The cyclic identity allows us to find the relation
\begin{align}
    F^{i_1\cdots i_k 0}=-kF^{0i_1\cdots i_k}.
\end{align}
Then the equation of motion is rewritten as
\begin{align}
    \del_{i_1}\del_{i_2}\cdots\del_{i_k} E^{i_1\cdots i_k}
=\rho,
\end{align}
where we have defined the charge density $\rho$ as
\begin{align}
    k \del_{i_1}\cdots\del_{i_{k-1}}J^{0i_1\cdots i_{k-1}}=\rho.
\end{align}
The equation of motion gives 
the current conservation $\del_{\mu_1} \cdots \del_{\mu_k} J^{\mu_1
\cdots \mu_k} = 0$ which implies
the relation,
\begin{align}
    \del_0 \rho + \del_{i_1\cdots i_k}J^{i_1\cdots i_k}=0,
\end{align}
where we have assumed $\partial_0\partial_0\partial_{i_3}\cdots\partial_{i_k}J^{00 i_3 \cdots i_k} = \cdots = \partial_0\cdots\partial_0\partial_{i_k}J^{000 \cdots
i_k} = \partial_0\cdots\partial_0J^{0 \cdots 0} = 0$ as in the case of the rank-2.
Then, we have the following conservation laws for the charges
$Q, Q^{j_1}, \ldots, Q^{j_1 \cdots j_{k-1}}$:
\begin{align}
    \begin{split}
    \del_0Q&=\del_0\int\! {\dop}V\,\rho=-\int\! {\dop}V\,\del_{i_1}\del_{i_2}\cdots\del_{i_k}J^{i_1\cdots i_k}=0,
    \\
    \del_0Q^{j_1}&=\del_0\int\! {\dop}V\,x^{j_1}\rho=-\int\! {\dop}V\,x^{j_1}\del_{i_1}\del_{i_2}\cdots\del_{i_k}J^{i_1\cdots i_k}=0,
    \\
    \vdots
    \\
    \del_0Q^{j_1\cdots j_{k-1}}&=\del_0\int\! {\dop}V\,x^{j_1}\cdots x^{j_{k-1}}\rho=-\int\! {\dop}V\,x^{j_1}\cdots x^{j_{k-1}}\del_{i_1}\del_{i_2}\cdots\del_{i_k}J^{i_1\cdots i_k}=0.
    \end{split}
\end{align}
These conservation laws forbid motions up to the multipoles of order $(k-2)$ 
while the $(k-1)$-th order multipoles are free to move.

The stress tensor $T_{\alpha\beta}$ is given by
\begin{align}
    T_{\alpha\beta}&=- \frac{2}{\sqrt{-g}}\frac{\delta }{\delta
 g^{\alpha\beta}}
\left. \int\! {\dop}^4x\,
\sqrt{-g} 
\left(
- \frac{1}{k (k+1)} 
F_{\mu_1 \cdots \mu_{k+1}} 
F^{\mu_1 \cdots \mu_{k+1}}
\right) \right|_{g^{\alpha\beta} = \eta^{\alpha\beta}}
\notag \\
&=-\frac{1}{k(k+1)}\eta_{\alpha\beta}F_{\mu_1\cdots\mu_{k+1}}F^{\mu_1\cdots\mu_{k+1}}+\frac{2}{k(k+1)}(kF_{\alpha}{}^{\mu_1\cdots\mu_k}F_{\beta\mu_1\cdots\mu_k}+F_{\mu_1\cdots\mu_k\alpha}F^{\mu_1\cdots\mu_k}{}_{\beta}).
\end{align}
The energy density $T_{00}$ is found to be
\begin{align}
    T_{00}=\frac{1}{k^2}(E_{i_1\cdots i_k}E^{i_1\cdots i_k}+B_{i_1}^{\:\:\:i_2\cdots i_k}B^{i_1}_{\:\:\:\:i_2\cdots i_k}).
\end{align}
This is again a generalization of the one for Maxwell theory.

\subsection{Hierarchy structure of symmetric gauge theories}
It is worthwhile to note that the field strength $F_{\mu \nu \rho}$ of
the symmetric gauge field of rank-2 is decomposed into the form:
\begin{align}
F_{\mu \nu \rho} 
=& \
\Big(
\del_{\mu} A_{\rho \nu} - \del_{\rho} A_{\mu \nu}
\Big)
+
\Big(
\del_{\nu} A_{\rho \mu} - \del_{\rho} A_{\nu \mu}
\Big)
\notag \\
=& \ 
F_{\mu \rho : \nu} + F_{\nu \rho : \mu},
\label{eq:bi-Maxwell}
\end{align}
where we have defined $F_{\mu \rho :\nu} = \del_{\mu} A_{\rho \nu} - \del_{\rho} A_{\mu \nu}$. This form is quite suggestive since $F_{\mu
\rho : \nu}$ is the field strength of the ``Maxwell field'' $A_{\mu}$
with one extra index $A_{\mu :\nu} = A_{\mu \nu}$.
Given the decomposition of the symmetric field strength $F_{\mu \nu
\rho}$ into the bi-Maxwell form \eqref{eq:bi-Maxwell}, the equation of
motion is 
\begin{align}
0 = \del_{\rho} F^{\mu \nu \rho}
= 
\del_{\rho} F^{\mu \rho : \nu}
+
\del_{\rho} F^{\nu \rho : \mu}.
\end{align}
We find that a solution to the Maxwell equation
\begin{align}
\del_{\rho} F^{\mu \rho : \nu} = 0
\label{eq:bi-Maxwell_eq}
\end{align}
gives a solution to the equation $\del_{\rho} F^{\mu \rho \nu} = 0$.
We stress that the extra index $\nu$ in the equation
\eqref{eq:bi-Maxwell_eq} is irrelevant to the structure of the differential
equations. Namely, it behaves like an internal index.

We then extend the above discussion to the higher rank symmetric gauge theories.
The field strength of the symmetric gauge field of rank-3 takes the form \eqref{eq:rank-r_field_strength}, 
which can be decomposed as 
\begin{align}
F_{\mu\nu\rho\sigma} 
&= \partial_\mu A_{\nu\rho\sigma} + \partial_\nu A_{\mu\rho\sigma} + \partial_\rho A_{\mu\nu\sigma} - 3 \partial_\sigma A_{\mu\nu\rho}
\notag \\
&= F_{\mu\sigma:\nu\rho} + F_{\nu\sigma:\mu\rho} + F_{\rho\sigma:\mu\nu}. 
\label{eq:rank-3_tri-Maxwell_form}
\end{align}
Here, we have defined $F_{\mu\sigma:\nu\rho} = \partial_\mu
A_{\sigma\nu\rho} - \partial_\sigma A_{\mu\nu\rho}$ 
as the ``Maxwell field strength'' with two extra indices, as we did for
the rank-2.
Similar to the rank-2 case, the field strength of the symmetric
gauge field of rank-3 can be rewritten in the ``tri-Maxwell'' form.
At the same time, the field strength of the rank-3 symmetric gauge field
can also be expressed as 
\begin{align}
F_{\mu\nu\rho\sigma} 
&= - \Big( \hat{F}_{\mu\sigma\nu:\rho} + \hat{F}_{\nu\sigma\rho:\mu} +
 \hat{F}_{\rho\sigma\mu:\nu} \Big), 
\label{eq:tri-rank-2_form}
\end{align}
where we have defined the strength of the rank-2 symmetric gauge field
with one extra index;
$\hat{F}_{\mu\nu\rho:\sigma} = \partial_\mu
A_{\nu\rho\sigma} + \partial_\nu A_{\mu\rho\sigma} - 2 \partial_\rho
A_{\mu\nu\sigma}$.
This is the ``tri-rank-2'' form of the rank-3 symmetric gauge field
strength.
Furthermore, the field strength $\hat{F}_{\mu\nu\rho:\sigma}$ in
\eqref{eq:tri-rank-2_form} is also decomposed into the bi-Maxwell form: 
\begin{align}
\hat{F}_{\mu\nu\rho:\sigma} 
&= F_{\mu\rho:\nu\sigma} + F_{\nu\rho:\mu\sigma}.
\end{align}
Substituting this into \eqref{eq:tri-rank-2_form} and using the following partial cyclic identity
\begin{align}
0 &= F_{\mu\nu:\rho\sigma} + F_{\nu\rho:\mu\sigma} + F_{\rho\mu:\nu\sigma},
\end{align}
we obtain the same tri-Maxwell form as in \eqref{eq:rank-3_tri-Maxwell_form}.
Thus, we find that the tri-Maxwell and the tri-rank-2 forms are equivalent. 

We now continue the discussion to the rank-4 case.
As in the case of the rank-3, the field strength of the symmetric gauge
field of rank-4 can be decomposed as
\begin{align}
F_{\mu\nu\rho\sigma\lambda} 
&= \partial_\mu A_{\nu\rho\sigma\lambda} + \partial_\nu A_{\mu\rho\sigma\lambda} 
	+ \partial_\rho A_{\mu\nu\sigma\lambda} + \partial_\sigma A_{\mu\nu\rho\lambda} 
	- 4 \partial_\lambda A_{\mu\nu\rho\sigma}
\notag \\
&= F_{\mu\lambda:\nu\rho\sigma} + F_{\nu\lambda:\mu\rho\sigma} + F_{\rho\lambda:\mu\nu\sigma} + F_{\sigma\lambda:\mu\nu\rho},
\label{eq:rank-4_quadra-Maxwell}
\end{align}
where we have defined the Maxwell field strength $F_{\mu\lambda:\nu\rho\sigma} = \partial_\mu A_{\lambda\nu\rho\sigma} - \partial_\lambda A_{\mu\nu\rho\sigma}$ 
with three extra indices.
This is the quadra-Maxwell form.
We also obtain the quadra-rank-3 form of the rank-4 symmetric gauge field strength 
\begin{align}
F_{\mu\nu\rho\sigma\lambda} 
&= - \Big( \hat{\hat{F}}_{\mu\nu\lambda\rho:\sigma}
	+ \hat{\hat{F}}_{\nu\rho\lambda\sigma:\mu}
	+ \hat{\hat{F}}_{\rho\sigma\lambda\mu:\nu}
	+ \hat{\hat{F}}_{\sigma\mu\lambda\nu:\rho} \Big)
\label{eq:rank-4_quadra-rank-3_form}
\end{align}
by using the rank-3 symmetric gauge field strength with one extra index: 
\begin{align}
\hat{\hat{F}}_{\mu\nu\rho\sigma:\lambda}
&= \partial_\mu A_{\nu\rho\sigma\lambda} + \partial_\nu A_{\mu\rho\sigma\lambda}
	+ \partial_\rho A_{\mu\nu\sigma\lambda} - 3 \partial_\sigma A_{\mu\nu\rho\lambda}.
\label{eq:rank3_with_extra_index}
\end{align}
The tensor \eqref{eq:rank3_with_extra_index} can be expressed as 
\begin{align}
\hat{\hat{F}}_{\mu\nu\rho\sigma:\lambda} 
&= F_{\mu\sigma:\nu\rho\lambda} + F_{\nu\sigma:\mu\rho\lambda} + F_{\rho\sigma:\mu\nu\lambda}
\notag \\
&= - \Big( \hat{F}_{\mu\sigma\nu:\rho\lambda} + \hat{F}_{\nu\sigma\rho:\mu\lambda} + \hat{F}_{\rho\sigma\mu:\nu\lambda} \Big)
\label{eq:rank-3_with_extra_index_alter}
\end{align}
by using $F_{\mu\nu:\rho\sigma\lambda}$ and the rank-2 symmetric gauge field strength with two extra indices 
$\hat{F}_{\mu\nu\rho:\sigma\lambda} 
= \partial_\mu A_{\nu\rho\sigma\lambda} + \partial_\nu A_{\mu\rho\sigma\lambda} - 2 \partial_\rho A_{\mu\nu\sigma\lambda}$. 
By substituting the first line of \eqref{eq:rank-3_with_extra_index_alter}
into \eqref{eq:rank-4_quadra-rank-3_form}, 
we obtain $F_{\mu\nu\rho\sigma\lambda}$ consisting of 12 Maxwell terms. 
Here, using the partial cyclic identity
\begin{align}
0 &= F_{\mu\nu:\rho\sigma\lambda} + F_{\nu\rho:\sigma\mu\lambda} 
	+ F_{\rho\sigma:\mu\nu\lambda} + F_{\sigma\mu:\nu\rho\lambda}, 
\end{align}
we can show that the result is equal to that of \eqref{eq:rank-4_quadra-Maxwell}.
On the other hand, substituting the second line in \eqref{eq:rank-3_with_extra_index_alter}
into \eqref{eq:rank-4_quadra-rank-3_form}, 
we obtain $F_{\mu\nu\rho\sigma\lambda}$ consisting of 12 ``rank-2''
terms $\hat{F}_{\mu\nu\rho:\sigma\lambda}$.
The following identities 
\begin{align}
0 &= \hat{F}_{\mu\rho\nu:\sigma\lambda} + \hat{F}_{\mu\rho\sigma:\nu\lambda} 
	+ \hat{F}_{\nu\sigma\rho:\mu\lambda} + \hat{F}_{\nu\sigma\mu:\rho\lambda}, 
\notag \\
0 &= \hat{F}_{\nu\rho\lambda:\mu\sigma} + \hat{F}_{\lambda\rho\mu:\nu\sigma} 
	+ \hat{F}_{\lambda\sigma\nu:\rho\mu} + \hat{F}_{\sigma\mu\lambda:\rho\nu} 
	+ \hat{F}_{\lambda\mu\rho:\sigma\nu} + \hat{F}_{\lambda\nu\sigma:\mu\rho}
\end{align}
yield the bi-rank-2 form: 
\begin{align}
F_{\mu\nu\rho\sigma\lambda} 
&= \hat{F}_{\mu\nu\lambda:\rho\sigma} + \hat{F}_{\rho\sigma\lambda:\mu\nu}.
\end{align}
Therefore, we again find a hierarchical structure in the field strength
of the symmetric gauge field of rank-4.
It is expressed in terms of the Maxwell, the rank-2, and also the rank-3
field strengths, respectively. 
These observations allow us to find an interesting hierarchical
structure of the field strengths for the symmetric gauge fields.
The several equivalent decomposability presented above allows us to
examine the field strength from various viewpoints.

In general, the rank-$k$ symmetric gauge field strength \eqref{eq:rank-r_field_strength}
can be decomposed into the ``multi-Maxwell'' form, 
consisting of $k$ terms of the Maxwell field strength 
$F_{\mu_1 \mu_{k+1}:\mu_2 \cdots \mu_k} 
= \partial_{\mu_1} A_{\mu_{k+1} \mu_2 \cdots \mu_k} - \partial_{\mu_{k+1}} A_{\mu_1 \mu_2 \cdots \mu_k}$
with $(k-1)$ extra indices. 
It can also be expressed in terms of rank-$(k-1)$, 
and the above discussion can be applied in the same way. 

The above discussion becomes a bit clearer when the symmetric gauge fields are 
expressed in the language of multi-forms. 
We begin by describing the relation between the rank-2 symmetric gauge field and bi-form. 
A {bi-form} of type $(p,q)$ is an element of $\Omega^{p,q}$~\cite{deMedeiros:2002qpr}.
Here $\Omega^{p,q} := \Omega^p \otimes \Omega^q$ is the $\group{GL}(D, \mathbb{R})$ irreducible tensor product 
of the space of $p$-forms $\Omega^p$ and the space of $q$-forms $\Omega^q$ on $\mathbb{R}^D$. 
In this language, the rank-2 symmetric gauge field can be expressed as 
$A^{(1,1)} = A_{\mu\nu} \, {\dop}x^\mu \otimes {\dop}x^\nu \in \Omega^{1,1}$. 
There are (left and right) exterior derivatives defined by $\LeftD : \Omega^{p,q} \to \Omega^{p+1,q}$ 
and $\RightD : \Omega^{p,q} \to \Omega^{p,q+1}$.  
Each of these exterior derivatives is nilpotent $\LeftD^2 = 0 = \RightD^2$ 
and commutative $\LeftD \RightD = \RightD \LeftD$. 
With these operators, the gauge transformation of $A^{(1,1)}$ is expressed as
\begin{align}
\delta A^{(1,1)} = 
\LeftD \RightD \lambda 
&= \partial_\mu \partial_\nu \lambda \, {\dop}x^\mu \otimes {\dop}x^\nu. 
\end{align}
Using the exterior derivatives, we define the field strength as $\mathbf{F} = (\LeftD - \RightD) A^{(1,1)}$. 
This field strength is shown to be gauge invariant from the nilpotency and commutativity of the exterior derivatives. 
The field strength $\mathbf{F}$ is the formal sum of a type $(2,1)$ bi-form and a type $(1,2)$ bi-form.
On the other hand, by rewriting the 2-form basis in terms of the type $(1,1)$ bi-form basis,
\begin{align}
{\dop}x^\mu \wedge {\dop}x^\nu 
&= \frac{1}{2} \big(
	{\dop}x^\mu \otimes {\dop}x^\nu 
	- {\dop}x^\nu \otimes {\dop}x^\mu \big), 
\label{eq:2-form_basis_and_bi-form_basis}
\end{align}
the field strength can be written in the type $(1,1,1)$ tri-form:
\begin{align}
\mathbf{F} 
&= (\LeftD - \RightD) A^{(1,1)} 
\notag \\
&= \frac{1}{2} \big( \partial_\mu A_{\nu\rho} - \partial_\nu A_{\mu\rho} \big)
	\, {\dop}x^\mu \wedge {\dop}x^\nu \otimes {\dop}x^\rho 
- \frac{1}{2} \big( \partial_\nu A_{\mu\rho} - \partial_\rho A_{\mu\nu} \big) 
	\, {\dop}x^\mu \otimes {\dop}x^\nu \wedge {\dop}x^\rho
\notag \\
&= \frac{1}{2} \big( \partial_\mu A_{\nu\rho} - 2 \partial_\nu A_{\mu\rho} + \partial_\rho A_{\mu\nu} \big) 
	\, {\dop}x^\mu \otimes {\dop}x^\nu \otimes {\dop}x^\rho
\notag \\
&= \frac{1}{2} F_{\mu\rho\nu} \, {\dop}x^\mu \otimes {\dop}x^\nu \otimes {\dop}x^\rho. 
\label{eq:bi-form_F_in_tri-form}
\end{align}
The tri-form is an element of the tensor product space of $\Omega^p$, $\Omega^q$, and $\Omega^r$. 
Typically, elements of the tensor product space of multiple $\Omega^{p_i}$ are referred to as {multi-forms}~\cite{deMedeiros:2002qpr}. 
The second equality in \eqref{eq:bi-form_F_in_tri-form} means that the
rank-2 symmetric gauge field can be written in the bi-Maxwell form. 
This discussion can be extended directly to the rank-$k$ symmetric gauge fields and multi-forms. 
It is shown explicitly that the field strength can be written in multi-Maxwell forms. 

We stress that the Bianchi identity for the bi-form field strength
$\mathbf{F}$ based on the exterior derivatives $\LeftD, \RightD$, 
is given by $(\LeftD + \RightD) \mathbf{F} = 0$.
This Bianchi identity is a formal sum of type $(3,1)$, type $(2,2)$, and type $(1,3)$ bi-forms. 
Written in components, the left-hand side of the Bianchi identity can be expressed as
\begin{align}
(\LeftD + \RightD) \mathbf{F} 
&= \frac{1}{2} \Big( 
\partial_\sigma F_{\mu\nu:\rho} \, 
	{\dop}x^\sigma \wedge {\dop}x^\mu \wedge {\dop}x^\nu \otimes {\dop}x^\rho 
+ \partial_\sigma F_{\mu\nu:\rho} \, 
	{\dop}x^\mu \wedge {\dop}x^\nu \otimes {\dop}x^\sigma \wedge {\dop}x^\rho \Big)
\notag \\
& \quad 
+ \frac{1}{2} \Big( 
\partial_\sigma F_{\nu\rho:\mu} \,
	{\dop}x^\sigma \wedge {\dop}x^\mu \otimes {\dop}x^\nu \wedge {\dop}x^\rho
+ \partial_\sigma F_{\nu\rho:\mu} \,
	+ {\dop}x^\mu \otimes {\dop}x^\sigma \wedge {\dop}x^\nu \wedge {\dop}x^\rho \Big),
\label{eq:bi-form_BI_in_bi-form_basis}
\end{align}
where $F_{\mu\nu:\rho} = \partial_\mu A_{\nu\rho} - \partial_\nu A_{\mu\rho}$ is the Maxwell field strength with one extra index. 
The Bianchi identity for $F_{\mu\nu:\rho}$ is given by $\LeftD^2 A^{(1,1)} = 0 = \RightD^2 A^{(1,1)}$ 
and expressed in terms of components as
\begin{align}
0 &= \partial_\mu F_{\nu\rho:\sigma} + \partial_\nu F_{\rho\mu:\sigma} + \partial_\rho F_{\mu\nu:\sigma}.
\label{eq:BI_for_Maxwell_with_one_extra_index}
\end{align}
Using this formula, \eqref{eq:bi-form_BI_in_bi-form_basis} can be rewritten as
\begin{align}
(\LeftD + \RightD) \mathbf{F} 
&= \frac{1}{2} \big( 
\partial_\rho F_{\mu\nu:\sigma} + \partial_\mu F_{\rho\sigma:\nu}
	\big) \, {\dop}x^\mu \wedge {\dop}x^\nu \otimes {\dop}x^\rho \wedge {\dop}x^\sigma
\notag \\
&= \frac{1}{4} \big( 
\partial_\rho F_{\mu\nu:\sigma} 
- \partial_\sigma F_{\mu\nu:\rho} 
+ \partial_\mu F_{\rho\sigma:\nu}
- \partial_\nu F_{\rho\sigma:\mu}
	\big) \, {\dop}x^\mu \otimes {\dop}x^\nu \otimes {\dop}x^\rho \otimes {\dop}x^\sigma.
\end{align}
Here, the second line is rewritten in type $(1,1,1,1)$ quadra-form using the basis relation \eqref{eq:2-form_basis_and_bi-form_basis}.
Using the formula \eqref{eq:BI_for_Maxwell_with_one_extra_index} again, we obtain 
\begin{align}
(\LeftD + \RightD) \mathbf{F} 
&= - \frac{1}{4} \big( 
\partial_\mu F_{\nu\rho:\sigma} 
+ \partial_\nu F_{\rho\mu:\sigma} 
- \partial_\mu F_{\nu\sigma:\rho} 
\notag \\
& \hspace{20mm}
- \partial_\nu F_{\sigma\mu:\rho} 
+ \partial_\mu F_{\rho\sigma:\nu}
- \partial_\nu F_{\rho\sigma:\mu}
	\big) \, {\dop}x^\mu \otimes {\dop}x^\nu \otimes {\dop}x^\rho \otimes {\dop}x^\sigma
\notag \\
&= \frac{1}{4} \Big[ 
\partial_\mu \big( 
	F_{\rho\nu:\sigma} + F_{\nu\sigma:\rho} + F_{\sigma\rho:\nu} \big)
\notag \\
& \hspace{20mm}
+ \partial_\nu \big( F_{\sigma\mu:\rho} + F_{\mu\rho:\sigma} + F_{\rho\sigma:\mu} \big)
	\Big] \, {\dop}x^\mu \otimes {\dop}x^\nu \otimes {\dop}x^\rho \otimes {\dop}x^\sigma.
\label{eq:BI_in_quadra-form}
\end{align}
On the flip side, 
the ``Bianchi identity'' $\varepsilon^{\alpha\mu\nu\rho} \partial_\mu F_{\beta\nu\rho} = 0$
can be rewritten as 
\begin{align}
0 &= \partial_{\mu} F_{\beta\nu\rho} 
	- \partial_{\mu} F_{\beta\rho\nu} 
	+  \partial_{\rho} F_{\beta\mu\nu} 
	- \partial_{\nu} F_{\beta\mu\rho} 
	+ \partial_{\nu} F_{\beta\rho\mu} 
	- \partial_{\rho} F_{\beta\nu\mu}
\notag \\
&= \partial_\beta \big( F_{\mu\rho:\nu} + F_{\nu\mu:\rho} + F_{\rho\nu:\mu} \big) 
+ 2 \big( \partial_\mu F_{\nu\rho:\beta} + \partial_\rho F_{\mu\nu:\beta} + \partial_\nu F_{\rho\mu:\beta} \big)
\notag \\
&= \partial_\beta \big( F_{\mu\rho:\nu} + F_{\nu\mu:\rho} + F_{\rho\nu:\mu} \big), 
\label{eq:fracton_BI_in_terms_of_Maxwell}
\end{align}
where we have used $F_{\mu\nu\rho} = F_{\mu\rho:\nu} + F_{\nu\rho:\mu}$ and \eqref{eq:BI_for_Maxwell_with_one_extra_index}.
Comparing \eqref{eq:BI_in_quadra-form} and \eqref{eq:fracton_BI_in_terms_of_Maxwell}, 
the latter is a sufficient condition of the former.
Thus, the ``Bianchi identity'' in
\eqref{eq:Bianchi_identity_for_rank-2_4d}
is a stronger version of the Bianchi identity $(\LeftD + \RightD) \mathbf{F} = 0$.
This is why we have called it with ``\quad''.
In the following, we employ the ``Bianchi identity''\footnote{
In the following, we omit ``\quad'' unless there is confusion.
} in
\eqref{eq:Bianchi_identity_for_rank-2_4d} rather than $(\LeftD +
\RightD) \mathbf{F} = 0$ and discuss the duality.

\section{Electric-magnetic duality for symmetric gauge theories} \label{sec:EM_dual}
We first start with the well-known electric-magnetic duality in the Maxwell
theory in four dimensions.
The electric-magnetic duality exchanges the equation of motion and the
Bianchi identity. This is trivially implemented by the properties of
differential forms.
The gauge equation of motion and the Bianchi identity
of the Maxwell field (1-form) $A = A_{\mu} \, {\dop}x^{\mu}$ in four spacetime dimensions are given by
\begin{align}
{\dop} \! *_4 \! F = 0, \qquad {\dop}F = 0,
\end{align}
where $F = {\dop}A$ is the field strength of the gauge field $A$ and 
$*_4$ is the Hodge star operator in four dimensions.
In the action level, we start from 
\begin{align}
S = \frac{1}{2} \int \! F \wedge *_4 F + \int \tilde{A} \wedge {\dop}F,
\label{eq:first_order_action_maxwell}
\end{align}
where the 2-form $F$ and the 1-form 
$\tilde{A}$ are independent fields.
Now we integrate out $\tilde{A}$.
The field equation is given by
\begin{align}
{\dop}F = 0.
\end{align}
This is the same form with the Bianchi identify for $A$ and we can
solve this equation by a 1-form $A$ as $F = {\dop}A$.
Note that this is the result of the Poincar\'e lemma in differential calculus.
Plugging this solution back into the action
\eqref{eq:first_order_action_maxwell}, we obtain the Maxwell action
\begin{align}
S = \frac{1}{2} \int \! F \wedge *_4 F, \qquad F = {\dop}A.
\end{align}
The 1-form $A$ is identified with the Maxwell field.

On the other hand, if we instead integrate out $F$, we have the
equation of motion
\begin{align}
F = - *_4 \! {\dop} \tilde{A} = - *_4 \! \tilde{F}.
\label{eq:Maxwell_dual}
\end{align}
Here we have defined $\tilde{F} = {\dop}\tilde{A}$.
By plugging the solution \eqref{eq:Maxwell_dual} back into the
action \eqref{eq:first_order_action_maxwell}, we obtain the dual action:
\begin{align}
S = \frac{1}{2} \int \! \tilde{F} \wedge *_4 \tilde{F}.
\end{align}
This is the magnetic dual of the original theory.
Note that the dual theory is apparently invariant under the gauge
transformation $\delta \tilde{A} = {\dop} \tilde{\lambda}$.
Here $\tilde{\lambda}$ is the gauge parameter.
One finds that the equation of motion and the Bianchi identity are
interchanged in the dual side:
\begin{align}
{\dop} \! *_4 \! F = 0, \quad {\dop} F = 0
\qquad 
\longleftrightarrow
\qquad
{\dop} \tilde{F} = 0, \quad {\dop} \! *_4 \! \tilde{F} = 0.
\end{align}
It is now almost obvious that the electric (magnetic) field in the
original theory corresponds to the magnetic (electric) field in the dual side.
The procedure is easily applied to the $p$-form electrodynamics.
The dual of the $p$-form theory in $D$-dimensional spacetime is the
theory of $(D-p-2)$-form in general.

\subsection{Symmetric gauge theory of rank-2}

It is natural to generalize the discussion to the symmetric gauge
theories. 
As we have shown, the electric-magnetic duality is defined by the exchange of the equation
of motion and the Bianchi identity.
For the symmetric gauge theory, we have the Bianchi identity\footnote{
As we have clarified, this is an identity but this does not have the same differential
geometrical underpinnings as in the case of differential $p$-forms.
}
\begin{align}
\varepsilon^{\alpha \nu \rho \sigma} \del_{\nu} F_{\rho \beta \sigma} =
 0,
\label{eq:rank2_Bianchi}
\end{align}
where $F_{\mu \nu \rho} = \del_{\mu} A_{\nu \rho} + \del_{\nu} A_{\mu
\rho} - 2 \del_{\rho} A_{\mu \nu}$ is the field strength of the
symmetric gauge field $A_{\mu \nu}$.
Based on this fact, we now consider the electric-magnetic dual theory of
the symmetric gauge field.
Following the case of the Maxwell theory, we start from the Lagrangian
\begin{align}
\mathcal{L} (F, \tilde{A}) 
= 
- \frac{1}{6} F_{\mu \nu \rho} F^{\mu \nu \rho}
+ \frac{2}{3} \tilde{A}^{\alpha \beta} \varepsilon_{\alpha} {}^{\mu \nu \rho} \del_{\mu}
 F_{\nu \beta \rho},
\label{eq:first_order_Lagrangian_rank2}
\end{align}
where $F_{\mu \nu \rho} = F_{\nu \mu \rho}$ and $\tilde{A}_{\mu \nu} =
\tilde{A}_{\nu \mu}$ are independent fields. 
Integrating out the field $\tilde{A}_{\alpha \beta}$ gives
\begin{align}
\varepsilon^{\alpha \mu \nu \rho} \del_{\mu} F_{\nu \beta \rho} = 0.
\end{align}
Since this is the same form of the Bianchi identity
\eqref{eq:rank2_Bianchi}, we have a solution
\begin{align}
F_{\mu \nu \rho} = \del_{\mu} A_{\nu \rho} + \del_{\nu} A_{\mu \rho} 
- 2 \del_{\rho} A_{\mu \nu},
\label{eq:rank2_Bianchi_solution}
\end{align}
where $A_{\mu \nu}$ is a symmetric gauge field.
Plugging the solution \eqref{eq:rank2_Bianchi_solution} back into the
Lagrangian \eqref{eq:first_order_Lagrangian_rank2}, we have the theory
of symmetric gauge field:
\begin{align}
\mathcal{L} (A) = - \frac{1}{6} F_{\mu \nu \rho} F^{\mu \nu \rho},
\qquad 
F_{\mu \nu \rho} = \del_{\mu} A_{\nu \rho} + \del_{\nu} A_{\mu \rho} 
- 2 \del_{\rho} A_{\mu \nu}.
\label{eq:original_Lagrangian}
\end{align}
On the other hand, if we integrate out $F_{\mu \nu \rho}$ in
\eqref{eq:first_order_Lagrangian_rank2}, the equation
of motion is given by
\begin{align}
F_{\mu \nu \rho} = \frac{1}{3} 
\Big(
\varepsilon_{\alpha \beta \nu \rho} \tilde{F}_{\mu} {}^{\alpha \beta}
+
\varepsilon_{\alpha \beta \mu \rho} \tilde{F}_{\nu} {}^{\alpha \beta}
\Big),
\label{eq:dual_relation}
\end{align}
where we have defined $\tilde{F}_{\mu \nu \rho} = \del_{\mu}
\tilde{A}_{\nu \rho} + \del_{\nu} \tilde{A}_{\mu \rho} - 2 \del_{\rho}
\tilde{A}_{\mu \nu}$.
Substituting this back into the Lagrangian
\eqref{eq:first_order_Lagrangian_rank2}, we find the dual theory
\begin{align}
\mathcal{L} (\tilde{A}) = 
- \frac{1}{6} \tilde{F}_{\mu \nu \rho} \tilde{F}^{\mu \nu \rho} +
 \frac{1}{12} \tilde{F}^{\alpha} {}_{\alpha \rho} \tilde{F}^{\beta}
 {}_{\beta} {}^{\rho},
\qquad
\tilde{F}_{\mu \nu \rho} = 
\del_{\mu} \tilde{A}_{\nu \rho} 
+ \del_{\nu} \tilde{A}_{\mu \rho} 
- 2 \del_{\rho} \tilde{A}_{\mu \nu}.
\label{eq:dual_Lagrangian}
\end{align}
Note that there appears a trace part of the field strength
$\tilde{F}^{\alpha} {}_{\alpha \mu}$ which was absent in $\mathcal{L} (A)$.
The dual action is invariant under the gauge transformation
$\delta \tilde{A}_{\mu \nu} = \del_{\mu} \del_{\nu} \tilde{\lambda}$.
We stress that the procedure is not apparent since there are no
mathematical structures that guarantees the dual side is written in the
manifestly gauge invariant field strength.
The equation of motion for the dual theory is 
\begin{align}
\del_{\rho} \tilde{F}^{\mu \nu \rho} 
+ \frac{1}{6}
\Big(
\del^{\mu} \tilde{F}^{\alpha} {}_{\alpha} {}^{\nu}
+
\del^{\nu} \tilde{F}^{\alpha} {}_{\alpha} {}^{\mu}
-
2 \eta^{\mu \nu} \del_{\rho} \tilde{F}^{\alpha} {}_{\alpha} {}^{\rho}
\Big)
= 0.
\end{align}
The relation \eqref{eq:dual_relation} implies that the electric
(magnetic) field in the original theory corresponds to the magnetic
(electric) field and each other.\footnote{
A democratic formulation based on the electric and the magnetic
potentials of the symmetric Maxwell theory is discussed in \cite{Bertolini:2022ijb}.
See also \cite{Avetisyan:2021heg} for a duality invariant formulation of gauge theories. }

It is useful to confirm the duality presented here actually exchanges
the equation of motion and the Bianchi identity.
Indeed, from the relation \eqref{eq:dual_relation}, we have
\begin{align}
\del^{\rho} F_{\mu \nu \rho} = 
\frac{1}{3}
\Big(
\varepsilon_{\alpha \beta \nu \rho} \del^{\rho} \tilde{F}_{\mu}
 {}^{\alpha \beta}
+
\varepsilon_{\alpha \beta \mu \rho} \del^{\rho} \tilde{F}_{\nu}
 {}^{\alpha \beta}
\Big).
\label{eq:dual_relation-2}
\end{align}
The $\text{LHS of } \eqref{eq:dual_relation-2} = 0$ is the equation
motion of the original theory while $\text{RHS of }
\eqref{eq:dual_relation-2} = 0$ gives the Bianchi identity for the dual
theory.
By the relation \eqref{eq:dual_relation}, we also have
\begin{align}
\frac{1}{3}
\Big(
\varepsilon^{\kappa \sigma \nu \rho} \del_{\sigma} F^{\mu} {}_{\nu \rho}
+
\varepsilon^{\mu \sigma \nu \rho} \del_{\sigma} F^{\kappa} {}_{\nu \rho}
\Big)
=
-
\left\{
\del_{\alpha} \tilde{F}^{\mu \kappa \alpha}
+
\frac{1}{6}
\Big(
\del^{\mu} \tilde{F}^{\alpha} {}_{\alpha} {}^{\kappa}
+
\del^{\kappa} \tilde{F}^{\alpha} {}_{\alpha} {}^{\mu}
-
2 \eta^{\kappa \mu} \del_{\alpha} \tilde{F}^{\beta} {}_{\beta} {}^{\alpha}
\Big)
\right\}.
\label{eq:dual_relation-3}
\end{align}
The $\text{LHS of } \eqref{eq:dual_relation-3} = 0$ gives the Bianchi
identity for the original theory while $\text{RHS of
}\eqref{eq:dual_relation-3} = 0$ 
gives the equation of motion for the dual theory.
Therefore the Lagrangian \eqref{eq:first_order_Lagrangian_rank2} precisely provides the
electric-magnetic dual theories.

We note that the situation of the symmetric gauge theory is quite
different from that of the ordinary Maxwell theory in which the
dual theory has the same form of the Lagrangian.
This is because the procedure is not based on the Hodge dual operation
of differential forms. In that case, the square of the Hodge operator is
proportional to the identity operator $*^2 = \pm 1$.
On the other hand, in our dual relation \eqref{eq:dual_relation}, 
the dual of the dual gives no identity. Namely, we have
\begin{align}
F_{\mu \nu \rho} 
\
\xrightarrow{\text{dual}} 
&\
\tilde{F}_{\mu \nu \rho} = 
\frac{1}{3} 
\Big(
\varepsilon_{\alpha \beta \nu \rho} F_{\mu} {}^{\alpha \beta}
+
\varepsilon_{\alpha \beta \mu \rho} F_{\nu} {}^{\alpha \beta}
\Big)
\notag \\ 
\
\xrightarrow{\text{dual}} 
&\
\frac{1}{3^2}
\varepsilon_{\alpha \beta \nu \rho}
\Big(
\varepsilon_{\alpha' \beta'} {}^{\alpha \beta} F_{\mu} {}^{\alpha'
 \beta'}
+
\varepsilon_{\alpha' \beta' \mu} {}^{\beta} F_{\alpha} {}^{\alpha' \beta'}
\Big)
\notag \\
&=
-
\Big(
F_{\mu \nu \rho}
+
\frac{1}{6} F^{\alpha} {}_{\alpha \nu} \eta_{\mu \rho}
+
\frac{1}{6} F^{\alpha} {}_{\alpha \mu} \eta_{\nu \rho}
-
\frac{1}{3} F^{\alpha} {}_{\alpha \rho} \eta_{\mu \nu} 
\Big).
\end{align}
Therefore the square of the dual operation in the symmetric theory is
not the identity.
This is why the dual Lagrangian $\mathcal{L} (\tilde{A})$ is not the
same form of the original one $\mathcal{L} (A)$.
We stress that although the dual Lagrangian is not the same form with
the original one, this procedure certainly exchanges the equation of motion and
the Bianchi identity and it provides the genuine electric-magnetic duality.
The characteristic feature of the dual operation also effects on the
BPS conditions which will be discussed later.

By introducing the coupling to a current $\tilde{J}^{\mu \nu}$, the equation of motion of the
dual theory is given by
\begin{align}
\del_{\rho} \tilde{F}^{\mu \nu \rho} 
+ \frac{1}{6}
\Big(
\del^{\mu} \tilde{F}^{\alpha} {}_{\alpha} {}^{\nu}
+
\del^{\nu} \tilde{F}^{\alpha} {}_{\alpha} {}^{\mu}
-
2 \eta^{\mu \nu} \del_{\rho} \tilde{F}^{\alpha} {}_{\alpha} {}^{\rho}
\Big)
=
\tilde{J}^{\mu \nu}
\end{align}
Then the current conservation follows trivially:
\begin{align}
\del_{\mu} \del_{\nu} \tilde{J}^{\mu \nu}
=
\del_{\mu} \del_{\nu} \del_{\rho} \tilde{F}^{\mu \nu \rho} 
+ \frac{1}{6}
\Big(
\del_{\mu} \del_{\nu}
\del^{\mu} \tilde{F}^{\alpha} {}_{\alpha} {}^{\nu}
+
\del_{\mu} \del_{\nu}
\del^{\nu} \tilde{F}^{\alpha} {}_{\alpha} {}^{\mu}
-
2 
\del_{\mu} \del_{\nu}
\eta^{\mu \nu} \del_{\rho} \tilde{F}^{\alpha} {}_{\alpha} {}^{\rho}
\Big)
=
0.
\end{align}
Here we have used the identity $\del_{\mu} \del_{\nu} \del_{\rho}
\tilde{F}^{\mu \nu \rho} = 0$.
Therefore the fractonic nature inherit also in this dual side.

\subsection{Duality for higher rank gauge theories}

We next examine the dual of the higher rank symmetric gauge theories in four dimensions. 
The above discussion can be extended directly to that of the symmetric
gauge field of rank-$k$.
The Bianchi identity for the symmetric gauge field of rank-$k$ is given by
\begin{align}
\varepsilon^{\alpha\mu\nu\rho} \partial_\mu F_{\beta_1 \cdots \beta_{k-1} \nu\rho} &= 0,
\end{align}
where $F_{\mu_1 \cdots \mu_{k+1}}$ is the field strength of the
symmetric gauge field of rank-$k$ defined in \eqref{eq:rank-r_field_strength}. 
Based on this Bianchi identity, we start from the following Lagrangian: 
\begin{align}
\mathcal{L} (F, \tilde{A})
&= - \frac{1}{k(k+1)} F_{\mu_1 \cdots \mu_{k+1}} F^{\mu_1 \cdots \mu_{k+1}} 
	+ 2c_k\, \tilde{A}^{\alpha\beta_1 \cdots \beta_{k-1}} \varepsilon_\alpha{}^{\mu\nu\rho} \partial_\mu F_{\beta_1 \cdots \beta_{k-1} \nu\rho},
\label{eq:rank-k_Lagrangian_with_aux_rank-k}
\end{align}
where $c_k = \sqrt{2/(3k(k+1))}$ is a constant.
The fields $F_{\mu_1 \cdots \mu_{k+1}} = F_{(\mu_1 \cdots \mu_{k})
\mu_{k+1}}$ and $\tilde{A}_{\mu_1 \cdots \mu_k} = \tilde{A}_{(\mu_1
\cdots \mu_k)}$ are independent.\footnote{
The symbols $(\mu_1 \cdots \mu_k)$ and $[\mu_1 \cdots \mu_k]$ 
indicate symmetrization and antisymmetrization of the indices, respectively.
}
If we integrate out $\tilde{A}_{\mu_1 \cdots \mu_k}$, we obtain the
Lagrangian \eqref{eq:Lagrangian_rank-k} for the symmetric gauge field of rank-$k$.
On the flip side, integrating out $F_{\mu_1 \cdots \mu_{k+1}}$ results
in the dual side.
The equation of motion for $F_{\mu_1 \cdots \mu_{k+1}}$ is given by
\begin{align}
F_{\beta_1 \cdots \beta_{k-2} \mu\nu\rho} 
&= - \frac{kc_k (k+1)}{2} \Big(
	\varepsilon^{\alpha_1 \alpha_2}{}_{\nu\rho} \partial_{\alpha_2} \tilde{A}_{\alpha_1 \beta_1 \cdots \beta_{k-2} \mu} 
	+ \varepsilon^{\alpha_1 \alpha_2}{}_{\mu\rho} \partial_{\alpha_2} \tilde{A}_{\alpha_1 \beta_1 \cdots \beta_{k-2} \nu} \Big)
\notag \\
&= \frac{kc_k}{2} \Big( 
	\varepsilon^{\alpha_1 \alpha_2}{}_{\nu\rho} \tilde{F}_{\alpha_1 \beta_1 \cdots \beta_{k-2} \mu \alpha_2} 
	+ \varepsilon^{\alpha_1 \alpha_2}{}_{\mu\rho} \tilde{F}_{\alpha_1 \beta_1 \cdots \beta_{k-2} \nu \alpha_2} \Big).
\label{eq:rel_F_and_tildeF_higher}
\end{align}
Here, $\tilde{F}_{\mu_1 \cdots \mu_{k+1}}$ is the field strength of $\tilde{A}_{\mu_1 \cdots \mu_k}$, defined by
\begin{align}
\tilde{F}_{\mu_1 \cdots \mu_{k+1}} 
&= \partial_{\mu_1} \tilde{A}_{\mu_2 \cdots \mu_{k+1}} 
	+ \partial_{\mu_2} \tilde{A}_{\mu_1 \mu_3 \cdots \mu_{k+1}} 
	+ \cdots 
	+ \partial_{\mu_k} \tilde{A}_{\mu_1 \cdots \mu_{k-1} \mu_{k+1}} 
	- k \partial_{\mu_{k+1}} \tilde{A}_{\mu_1 \cdots \mu_k}. 
\end{align}
Substituting the relation \eqref{eq:rel_F_and_tildeF_higher} into 
the original Lagrangian \eqref{eq:rank-k_Lagrangian_with_aux_rank-k} yields
the dual Lagrangian 
\begin{align}
\mathcal{L}(\tilde{A}) 
&= - \frac{kc_k^2}{2(k+1)} \bigg(
	\frac{3(k+1)}{k} \tilde{F}_{\mu_1 \cdots \mu_{k+1}} \tilde{F}^{\mu_1 \cdots \mu_{k+1}} 
	- \frac{k+7}{4} \tilde{F}_{\mu_1 \cdots \mu_{k-2} \rho}{}^{\rho\nu} \tilde{F}^{\mu_1 \cdots \mu_{k-2} \sigma}{}_{\sigma\nu}
\notag \\
& \hspace{35mm}
	- \frac{(k-2)(k+3)}{4} 
		\tilde{F}_{\mu_1 \cdots \mu_{k-3} \alpha \nu}{}^{\nu\beta} 
		\tilde{F}^{\mu_1 \cdots \mu_{k-3} \beta \rho}{}_{\rho\alpha} \bigg).
\label{eq:dual_Lagrangan_of_rank-k}
\end{align}
There appears the new trace part of the field strength 
that did not exist in the rank-2 case.
Indeed, the third term in $\mathcal{L}(\tilde{A})$ vanishes when $k=2$. 
The Lagrangian $\mathcal{L} (\tilde{A})$ is invariant under the gauge transformation
$\delta \tilde{A}_{\mu_1 \cdots \mu_k} = \partial_{\mu_1} \cdots \partial_{\mu_k} \tilde{\lambda}$. 
The relation \eqref{eq:rel_F_and_tildeF_higher} between $F_{\mu_1 \cdots \mu_{k+1}}$ and $\tilde{F}_{\mu_1 \cdots \mu_{k+1}}$
implies that the roles of the electric and the magnetic fields are inter\-changed
in the original and the dual theories. 

In the following, we show that the duality swaps the roles of 
the equation of motion and the Bianchi identity.
From the Lagrangian \eqref{eq:dual_Lagrangan_of_rank-k}, 
we find that the equation of motion for $\tilde{F}$ is given by 
\begin{align}
0 &= 6 \Big[ \partial_\nu \tilde{F}^{\mu_1 \cdots \mu_k\nu} - \partial_\nu \tilde{F}^{\nu (\mu_1 \cdots \mu_k)} \Big]
- (k-2) {\eta^{(\mu_{k-1}\mu_{k}|} \partial_{\nu} \tilde{F}^{\nu | \mu_1 \cdots \mu_{k-3} | \rho}{}_{\rho}{}^{|\mu_{k-2})}}
\notag \\
& \quad
+ (k+1) {\partial^{(\mu_1} \tilde{F}^{\mu_2 \cdots \mu_{k-2} \mu_{k-1} | \rho}{}_{\rho}{}^{|\mu_k)}}
- 3 {\eta^{(\mu_{k-1}\mu_{k}|} \partial_{\nu} \tilde{F}^{|\mu_1 \cdots \mu_{k-2}) \rho}{}_{\rho}{}^{\nu}} . 
\label{eq:rank-k_EoM_for_tildeF}
\end{align}
Using the equation \eqref{eq:rel_F_and_tildeF_higher}, 
we find that the Bianchi identity symmetrized by the uncontracted indices is
calculated as 
\begin{align}
\varepsilon_{\alpha_1 \alpha_2}{}^{(\mu_1| \rho} \partial_\rho F^{|\mu_2 \cdots \mu_k) \alpha_1 \alpha_2}
&= - \frac{kc_k}{4} \bigg[
	6 \Big( \partial_\rho \tilde{F}^{(\mu_1 \cdots \mu_k)\rho}
		- \partial_\rho \tilde{F}^{\rho(\mu_1 \cdots \mu_k)} \Big)
	+ (k+1) \partial^{(\mu_k|} \tilde{F}_{\alpha_1}{}^{\alpha_1 |\mu_1 \cdots \mu_{k-1})}
\notag \\
& \hspace{16mm}
	- (k-2) \eta^{(\mu_1 \mu_k|} \partial_\rho \tilde{F}^\rho{}_{\alpha_1}{}^{\alpha_1 |\mu_2 \cdots \mu_{k-1})}
	- 3 \eta^{(\mu_1 \mu_k|} \partial_\rho \tilde{F}_{\alpha_1}{}^{\alpha_1 |\mu_2 \cdots \mu_{k-1}) \rho}
	\bigg].  
\label{eq:BI_for_F_and_EoM_for_tildeF}
\end{align}
The left-hand side of \eqref{eq:BI_for_F_and_EoM_for_tildeF} vanishes by the Bianchi identity for $F$, 
and the right-hand side vanishes by the equation of motion for $\tilde{F}$ \eqref{eq:rank-k_EoM_for_tildeF}. 
On the other hand, substituting \eqref{eq:rel_F_and_tildeF_higher} into the equation of motion for $F$, 
we find  
\begin{align}
\partial^\rho F_{\beta_1 \cdots \beta_{k-2} \mu\nu\rho}
&= \frac{kc_k}{2} \Big( 
	\varepsilon^{\alpha_1 \alpha_2}{}_{\nu\rho} 
		\partial^\rho \tilde{F}_{\alpha_1 \beta_1 \cdots \beta_{k-2} \mu \alpha_2}
	+ \varepsilon^{\alpha_1 \alpha_2}{}_{\mu\rho} 
		\partial^\rho \tilde{F}_{\alpha_1 \beta_1 \cdots \beta_{k-2} \nu \alpha_2}\Big).
\label{eq:EoM_for_F_and_BI_for_tildeF}
\end{align}
The left-hand side of this equation vanishes due to the equation of motion for $F$, 
and the right-hand side vanishes due to the Bianchi identity for $\tilde{F}$. 
Thus, as in the rank-2 case, the rank-$k$ Lagrangian \eqref{eq:rank-k_Lagrangian_with_aux_rank-k}
provides the electric-magnetic dual theory.

\section{Dualities in diverse dimensions and mixed symmetric tensors}
\label{sec:duality_higher_dimensions}

\subsection{Dualities in higher dimensions}
In this section, we discuss the duality in higher dimensions greater
than four.
We start with the symmetric gauge field of rank-2.
The generalization to the rank-$k$ and the multipoles will be discussed later.

\paragraph{$D=5$.}
We consider the duality for the symmetric gauge fields of rank-2 in five dimensions. 
The Bianchi identity in five-dimensional spacetime is given by 
\begin{align}
\varepsilon^{\alpha_1 \alpha_2 \mu\nu\rho} 
\partial_\mu F_{\beta\nu\rho} 
&= 0 \qquad 
(\alpha_1, \alpha_2, \beta, \mu, \nu, \ldots = 0,1,\ldots, 4).
\end{align}
Following the prescription in the previous section, 
a Lagrangian $\mathcal{L} (F, \tilde{A})$ that we may start from contains
terms of the form $\tilde{A}^{\alpha_1 \alpha_2 | \beta}
\varepsilon_{\alpha_1 \alpha_2} {}^{\mu \nu \rho} \del_{\mu} F_{\beta
\nu \rho}$.
Here the symbol $\tilde{A}^{\mu \nu | \rho}$ means that the only first
two indices $\mu, \nu$ are antisymmetric, $\tilde{A}^{\mu \nu | \rho} =
- \tilde{A}^{\nu \mu | \rho}$.
This implies that the dual of the symmetric gauge field $A_{\mu \nu}$
is described by the mixed symmetric tensor field $\tilde{A}_{\mu \nu |
\rho}$. We call this the type $(2,1)$ field in the following.

We find that it is convenient to start from the mixed symmetric tensor
gauge theory rather than the symmetric rank-2 theory.
We thereby change the starting point in this section.
The mixed symmetric tensor gauge field ${A}_{\mu\nu | \rho}$ 
has the structure of a 2-form field with the one extra index. 
Hence, the gauge transformation of this field should be of the form:
\begin{align}
A_{\mu\nu|\rho} 
&\mapsto A_{\mu\nu|\rho} 
+ \partial_\rho \big( \partial_\mu \lambda_\nu - \partial_\nu
 \lambda_\mu \big), 
\label{eq:(2,1)_gauge_transf}
\end{align}
where $\lambda_\mu$ is the 1-form gauge parameter. 
Given the gauge transformation \eqref{eq:(2,1)_gauge_transf}, 
we find that the gauge invariant field strength is given by
\begin{align}
\mathcal{F}_{\mu\nu\rho\sigma}
&= 3! \Big[ \big( \partial_{[\mu} A_{\rho\sigma]|\nu} \big)
	+ \big( \partial_{[\nu} A_{\rho\sigma]|\mu} \big) 
	+ \big( \partial_{[\sigma} A_{\mu\nu]|\rho} \big) \Big].
\label{eq:field_strength_for_mixed(1,2)}
\end{align}
This field strength satisfies the ``Bianchi identity'' 
\begin{align}
\varepsilon^{\alpha\mu\nu\rho\sigma} 
\partial_\mu \mathcal{F}_{\beta\nu\rho\sigma} 
&= 0.
\label{eq:Bianchi_for_mixed(1,2)_in_5d}
\end{align}
Based on these facts, we have the following Lagrangian: 
\begin{align}
\mathcal{L} (\mathcal{F}, \tilde{A})
&= - \frac{1}{6} \mathcal{F}_{\mu\nu\rho\sigma} \mathcal{F}^{\mu\nu\rho\sigma} 
+ \frac{1}{3} \tilde{A}^{\alpha\beta} \varepsilon_{\alpha}{}^{\mu\nu\rho\sigma} 
	\partial_\mu \mathcal{F}_{\beta\nu\rho\sigma}, 
\label{eq:rank-2_sym_and_mixed_type_(1,2)_L2}
\end{align}
where $\mathcal{F}_{\mu\nu\rho\sigma}$ and $\tilde{A}^{\alpha\beta}$ are independent fields. 
If we integrate out $\tilde{A}_{\mu \nu}$, we obtain the equation that
is the same form of the Bianchi identity~\eqref{eq:Bianchi_for_mixed(1,2)_in_5d}.
It is obvious that the expression
\eqref{eq:field_strength_for_mixed(1,2)} is a solution to this equation.
Substituting the solution \eqref{eq:field_strength_for_mixed(1,2)} into
\eqref{eq:rank-2_sym_and_mixed_type_(1,2)_L2}, the Lagrangian becomes
\begin{align}
\mathcal{L} (A) 
&= - \frac{1}{6} \mathcal{F}_{\mu\nu\rho\sigma} \mathcal{F}^{\mu\nu\rho\sigma},
\end{align}
where $\mathcal{F}_{\mu \nu \rho \sigma}$ is defined in \eqref{eq:field_strength_for_mixed(1,2)}.
This is nothing but the Lagrangian of the mixed symmetric tensor gauge field ${A}_{\mu\nu | \rho}$.
On the other hand, if we integrate out $\mathcal{F}_{\mu \nu \rho \sigma}$, we obtain the following equation:
\begin{align}
\mathcal{F}_{\mu\nu\rho\sigma} 
&= - \varepsilon^{\alpha\beta}{}_{\nu\rho\sigma} \partial_\beta \tilde{A}_{\alpha\mu} 
= \frac{1}{3} \varepsilon^{\alpha\beta}{}_{\nu\rho\sigma} \tilde{F}_{\mu\alpha\beta},
\label{eq:rel_rank-2-sym_and_mixed(1,2)}
\end{align}
where $\tilde{F}_{\mu\alpha\beta} = 
\partial_\mu \tilde{A}_{\alpha\beta} 
+ 
\partial_\alpha \tilde{A}_{\mu\beta} 
- 2 \partial_\beta \tilde{A}_{\mu\alpha}$
is the field strength of $\tilde{A}_{\mu \nu}$.
Substituting the equation \eqref{eq:rel_rank-2-sym_and_mixed(1,2)} into~\eqref{eq:rank-2_sym_and_mixed_type_(1,2)_L2}, 
we obtain the dual Lagrangian
\begin{align}
\mathcal{L} (\tilde{A}) 
&= - \frac{1}{6} \tilde{F}_{\mu\nu\rho} \tilde{F}^{\mu\nu\rho}, 
\end{align}
where we have used the relation $\tilde{F}^{\mu\alpha\beta} \big(
\tilde{F}_{\mu\alpha\beta} + 2 \tilde{F}_{\mu\beta\alpha} \big) = 0$. 

For the type $(2,1)$ tensor field, 
by substituting the dual relation \eqref{eq:rel_rank-2-sym_and_mixed(1,2)} 
into a symmetric combination of the Bianchi 
identity \eqref{eq:Bianchi_for_mixed(1,2)_in_5d}, 
we obtain the equation of motion for the symmetric gauge field of rank-2:
\begin{align}
0 =& \ 
\varepsilon^{\alpha\mu\nu\rho\sigma}\del_{\mu}\mathcal{F}_{\beta\nu\rho\sigma}
+
\varepsilon_{\beta\mu\nu\rho\sigma}\del^{\mu}\mathcal{F}^{\alpha\nu\rho\sigma}
\notag \\
=& \
\del_{\mu}\tilde{F}_{\beta}{}^{\alpha\mu}-\del_{\mu}\tilde{F}_{\beta}{}^{\mu\alpha}+\del^{\mu}\tilde{F}^{\alpha}{}_{\beta\mu}-\del^{\mu}\tilde{F}^{\alpha}{}_{\mu\beta} 
\notag \\
=& \ 
3\del_{\mu} \tilde{F}^{\alpha} {}_{\beta} {}^{\mu}
\end{align}
Here we have used the relation 
$\tilde{F}_{\beta\alpha\mu}=-\tilde{F}_{\alpha\mu\beta}-\tilde{F}_{\mu\beta\alpha}$ 
that follows from the cyclic identity.
Note that we need to symmetrize the Bianchi identity to obtain the
equation of motion since the latter is symmetric with respect to the
uncontracted indices.
On the other hand, the equation of motion for the type $(2,1)$ field is
given by 
\begin{align}
0
&=
\del_{\sigma}
\Big(
\mathcal{F}^{\sigma\rho\mu\nu}
+\mathcal{F}^{\nu\rho\sigma\mu}
+\mathcal{F}^{\mu\rho\nu\sigma}
+\mathcal{F}^{\rho\sigma\mu\nu}
+\mathcal{F}^{\rho\nu\sigma\mu}
\notag \\
& \quad \qquad 
+\mathcal{F}^{\rho\mu\nu\sigma}
+\mathcal{F}^{\mu\nu\rho\sigma}
-\mathcal{F}^{\sigma\mu\rho\nu}
+\mathcal{F}^{\nu\sigma\rho\mu}
-(\mu\leftrightarrow\nu)\Big).
\end{align}
Substituting the relation \eqref{eq:rel_rank-2-sym_and_mixed(1,2)} into
the equation of motion, 
we obtain the Bianchi identity for the symmetric gauge field of rank-2 $\tilde{A}_{\mu\nu}$:
\begin{align}
    0=\varepsilon^{\alpha_1\alpha_2}{}^{\mu\nu\rho}\del_{\mu}\tilde{F}_{\beta_1\nu\rho}.
\end{align}
Thus, the equation of motion and the Bianchi identity for the type $(2,1)$ field corresponds to the
Bianchi identity and the equation of motion for the rank-2 symmetric
field, respectively.
Therefore, we conclude that the symmetric gauge field
$\tilde{A}_{\mu\nu}$ of rank-2 and the mixed symmetric tensor gauge
field $A_{\mu\nu|\rho}$ are dual in five dimensions.

We next consider the dual of the higher rank symmetric gauge field.
For the field strength $F_{\mu_1 \cdots \mu_{k+1}}$ of the symmetric
gauge field $A_{\mu_1 \cdots \mu_k}$ of rank-$k$, 
the Bianchi identity in five-dimensional spacetime is given by 
\begin{align}
\varepsilon^{\alpha_1 \alpha_2 \mu\nu\rho} \partial_\mu F_{\beta \cdots \beta_{k-1} \nu\rho} = 0.
\end{align}
If we make a Lorentz invariant term based on this Bianchi identity, 
then a mixed symmetric tensor field $\tilde{A}_{\alpha_1 \alpha_2 | \beta_1 \cdots \beta_{k-1}}$ 
(with two antisymmetric and $(k-1)$ symmetric indices) appears as an
independent field in the Lagrangian. 
As in the rank-2 symmetric theory, the duality can be shown by starting
from a Lagrangian containing the type $(2,k-1)$ field and
$\tilde{A}_{\mu \nu}$, and by integrating out either field.
This means that the mixed symmetric tensor field $A_{\mu_1 \mu_2 | \nu_1 \cdots \nu_{k-1}}$ 
naturally appears as the dual of the symmetric gauge field of rank-$k$ in five dimensions.

\paragraph{$D \ge 6$.}
The same discussion holds for dualities in six dimensions.
In the six-dimensional spacetime, we expect that the dual of the
symmetric tensor of rank-2 is the mixed symmetric tensor of type $(3,1)$.
The mixed symmetric tensor gauge field $A_{\mu\nu\rho|\sigma}$ is a 3-form
with one extra index. 
The gauge transformation based on the fracton gauge principle is defined
by 
\begin{align}
    A_{\mu\nu\rho|\sigma} 
&\mapsto A_{\mu\nu\rho|\sigma} 
+ \partial_\sigma \big( \partial_\mu \lambda_{\nu\rho} + \partial_\nu
 \lambda_{\rho\mu} +\partial_\rho \lambda_{\mu\nu} \big).
\label{eq:(3,1)_gauge_transf}
\end{align}
Here $\lambda_{\mu\nu} = - \lambda_{\nu \mu}$ is a 2-form gauge
parameter.
The field strength that is invariant under the gauge transformation
\eqref{eq:(3,1)_gauge_transf} is found to be 
\begin{align}
    \mathcal{F}_{\mu\nu\rho\sigma\kappa}= 4! \Big[ \big( \partial_{[\mu} A_{\rho\sigma\kappa]|\nu} \big)
	+ \big( \partial_{[\nu} A_{\rho\sigma\kappa]|\mu} \big) 
	+ \big( \partial_{[\kappa} A_{\mu\nu\rho]|\sigma} \big)\Big].
\label{eq:field_strength_for_mixed(1,3)}
\end{align}
Note that this field strength satisfies the Bianchi identity
\begin{align}
    \varepsilon^{\alpha\mu\nu\rho\sigma\kappa}\del_{\mu}\mathcal{F}_{\beta\nu\rho\sigma\kappa}=0.
\label{eq:Bianchi_for_mixed(1,3)_in_6d}
\end{align}
Based on these facts, we consider the following Lagrangian:
\begin{align}
    \mathcal{L}(\mathcal{F}, \tilde{A})
&= - \frac{1}{24} \mathcal{F}_{\mu\nu\rho\sigma\kappa} \mathcal{F}^{\mu\nu\rho\sigma\kappa} 
+ \frac{1}{12} \tilde{A}^{\alpha\beta} \varepsilon_{\alpha}{}^{\mu\nu\rho\sigma\kappa} 
	\partial_\mu \mathcal{F}_{\beta\nu\rho\sigma\kappa}, \label{eq:rank-2_sym_and_mixed_type_(1,3)_L2}
\end{align}
where $\mathcal{F}_{\mu\nu\rho\sigma\kappa}$ and
$\tilde{A}^{\alpha\beta}$ are independent fields.
Integrating out $\tilde{A}^{\alpha \beta}$ gives the equation that is the same form
with the Bianchi identity \eqref{eq:Bianchi_for_mixed(1,3)_in_6d}.
We can easily confirm that the expression
\eqref{eq:field_strength_for_mixed(1,3)} satisfies this equation.
Then the Lagrangian becomes that of the type $(3,1)$ tensor field:
\begin{align}
    \mathcal{L}(\mathcal{F}, \tilde{A})
&= - \frac{1}{24} \mathcal{F}_{\mu\nu\rho\sigma\kappa} \mathcal{F}^{\mu\nu\rho\sigma\kappa}.
\end{align}
On the other hand, integrating out the field $\mathcal{F}_{\mu \nu \rho
\sigma \kappa}$ gives the following equation
\begin{align}
\mathcal{F}_{\mu\nu\rho\sigma\kappa} 
&= - \varepsilon^{\alpha\beta}{}_{\nu\rho\sigma\kappa} \partial_\beta \tilde{A}_{\alpha\mu} 
= \frac{1}{3} \varepsilon^{\alpha\beta}{}_{\nu\rho\sigma\kappa} \tilde{F}_{\mu\alpha\beta}.
\label{eq:rel_rank-2-sym_and_mixed(1,3)}
\end{align}
Substituting this equation into
\eqref{eq:rank-2_sym_and_mixed_type_(1,3)_L2}, we obtain 
\begin{align}
\mathcal{L} (\tilde{A}) 
&= - \frac{1}{6} \tilde{F}_{\mu\nu\rho} \tilde{F}^{\mu\nu\rho}.
\end{align}
Here we have used the relation
$\tilde{F}^{\mu\alpha\beta}(\tilde{F}_{\mu\alpha\beta}+2\tilde{F}_{\mu\beta\alpha})=0$.
Therefore we have shown that the rank-2 symmetric tensor
$\tilde{A}_{\mu\nu}$ is dual to the mixed symmetric tensor 
$A_{\mu\nu\rho|\sigma}$ of type $(3,1)$ in six dimensions.

For the symmetric gauge field of rank-$k$, by utilizing the Bianchi identity
\begin{align}
\varepsilon^{\alpha_1 \alpha_2 \alpha_3 \mu\nu\rho} \partial_\mu
 F_{\beta_1 \cdots \beta_{k-1} \nu\rho} = 0,
\end{align}
we can perform the same procedure as in the five-dimensional case.
Therefore the mixed symmetric tensor field of type $(3,k-1)$ is dual to the
symmetric tensor gauge field of rank-$k$ in six dimensions.

When the spacetime dimension is greater than four, 
mixed symmetric tensor fields generically appear in the dual of the symmetric tensor gauge theories. 
In $D$ spacetime dimensions, the dual field of the symmetric gauge field
of rank-$k$ is the mixed symmetric tensor gauge field of type $(D-3,k-1)$.
This appears as $A_{\mu_1 \cdots \mu_{D-3}| \nu_1 \cdots \nu_{k-1}}$
where the first $(D-3)$ indices $\mu_1 \cdots \mu_{D-3}$ are totally
antisymmetric while the last $(k-1)$ indices $\nu_1 \cdots \nu_{k-1}$
are totally symmetric.
The symmetry of the indices in the mixed symmetric tensor fields is easily understood by the Young tableaux.
The mixed symmetric tensor of type $(2,1)$ is represented by 
{\scriptsize $\ydiagram[]{2,1}$} which has been known to be the Curtright field
\cite{Curtright:1980yk}. 
For each increase by one spacetime dimension, one $\ydiagram[]{1}$ 
in the first column of the mixed symmetric tensor field is increased by one. 
For one rank increase, there is one more $\ydiagram[]{1}$ in the row of
the mixed symmetric tensor field. 
The vertical column corresponds to the spacetime dimension while the horizontal row to the multipoles.
In four dimensions, there is only one box in the first column, thus the indices of the dual field are totally symmetric.
The mixed symmetric tensor fields in the dual side of the symmetric
gauge fields of rank-$k$ in diverse dimensions, are listed in
Table~\ref{tab:mixed_sym_tensor_and_Young_tableau}.
\begin{table}[t]
\centering
\begin{tabular}{| c | c | c |}
\hline 
\textbf{dim} & \textbf{gauge fields} & \textbf{dual fields and Young tableaux} \\
\hline 
4 & $A_{\mu_1 \cdots \mu_k}$ 
& $\tilde{A}_{\mu_1 \cdots \mu_k} \sim \overbrace{\ydiagram[]{8}}^{\text{$k$ boxes}}
\vphantom{\ydiagram[]{1,1,1,1,1}}$ \\
\hline 
5 & $A_{\mu_1 \cdots \mu_k}$ 
& $\tilde{A}_{\mu_1 \mu_2 | \nu_1 \cdots \nu_{k-1}} \sim \overbrace{\ydiagram[]{8,1}}^{\text{$k$ boxes}}
\vphantom{\ydiagram[]{1,1,1,1,1,1}}$ \\
\hline 
$\vdots$ & $\vdots$ & $\vdots$ \\
\hline 
$D$ & $A_{\mu_1 \cdots \mu_k}$ 
& $\tilde{A}_{\mu_1 \cdots \mu_{D-3} | \nu_1 \cdots \nu_{k-1}} 
\sim \text{\scriptsize $(D-3)$ boxes} \left\{ \vphantom{\ydiagram[]{8,1,1,1}} \right. \!\!
\overbrace{\ydiagram[]{8,1,1,1}}^{\text{$k$ boxes}} 
\vphantom{\ydiagram[]{1,1,1,1,1,1,1,1}}$ \\
\hline 
\end{tabular}
\caption{The mixed symmetric tensor gauge fields as duals of the
 symmetric gauge fields of rank-$k$ in each dimension.
The corresponding Young tableaux are shown.}
\label{tab:mixed_sym_tensor_and_Young_tableau}
\end{table}

\subsection{Dualities in lower dimensions}

\paragraph{$D=3$.}
In three-dimensional spacetime, only the electric-magnetic dual between 1-form and 0-form is allowed. 
The dual of higher rank symmetric gauge fields cannot be well-defined. 
When we consider the rank-$k$ symmetric gauge field $A_{\mu_1 \cdots \mu_k}$ ($\mu_i = 0,1,2$), 
the Bianchi identity can be written as $\varepsilon^{\mu\nu\rho} \partial_\mu F_{\alpha_1 \cdots \alpha_{k-1} \nu\rho} = 0$, 
hence it seems possible to construct the following Lagrangian 
\begin{align}
\mathcal{L} (F, \tilde{A}) 
&= - \frac{1}{k(k+1)} F_{\mu_1 \cdots \mu_{k+1}} F^{\mu_1 \cdots \mu_{k+1}} 
+ b \tilde{A}^{\alpha_1 \cdots \alpha_{k-1}} \varepsilon^{\mu\nu\rho} \partial_\mu F_{\alpha_1 \cdots \alpha_{k-1} \nu\rho} 
\label{eq:test_Lagrangian_in_3d}
\end{align}
where $b$ is a constant.
For this Lagrangian, integrating out $\tilde{A}$ yields the Lagrangian $\mathcal{L} (A)$.
However, if we integrate out $F$, the resulting Lagrangian cannot be
described in terms of the field strength of $\tilde{A}$. 
This is due to the fact that the indices of $\tilde{A}_{\alpha_1 \cdots
\alpha_{k-1}}$ and $\varepsilon^{\mu\nu\rho}$ are not contracted.  
Since the dual side is no longer gauge invariant, 
we see that the dual transformation using the Lagrangian \eqref{eq:test_Lagrangian_in_3d} is not possible except for $k=1$.

\paragraph{$D=2$.}
In two-dimensional spacetime, the duality between 0-forms is the only possible example. 
Even if we consider the higher rank symmetric gauge field, 
we cannot construct a Lagrangian with an auxiliary field 
because the Bianchi identity cannot be expressed using the totally
antisymmetric symbol $\varepsilon_{\mu\nu}$ ($\mu, \nu = 0,1$). 
Considering the 0-form to be a compact scalar field,
this is nothing but the T-duality of the target space in string theory
\cite{Buscher:1987sk, Buscher:1987qj}.

\section{Mixed symmetric tensor fields and immobile $p$-branes} \label{sec:immobile_p-branes}
In this section, we study physical consequences of the mixed symmetric
tensor fieds discussed in the previous section.
We found that there appear mixed symmetric tensor gauge fields in the dual of the
symmetric gauge field of rank-2 in $D \ge 5$.
We start by the mixed symmetric tensor field $A_{\mu \nu| \rho}$ of
the type $(2,1)$ Young tableau in $D=5$ as a prototypical example.
It has been shown that such a type $(2,1)$ field plays a role of the gauge field
\cite{Curtright:1980yk}.
Furthermore, it is well-known that a type $(2,1)$ field appears as a
dual of the graviton field $h_{\mu \nu}$ in $D=5$ \cite{Hull:2000zn}.
Indeed, the electric-magnetic dual of the linearized gravitational field
(a symmetric tensor of rank-2) in $D$ dimensions is described by the type $(D-3,1)$ mixed
symmetric tensor field which is analogous to our case.
Although the situation is similar to ours, we emphasize that the result
is significantly different.
The reason of difference comes from the gauge symmetry.
The gauge transformation of the linearized gravitational (graviton) field
$h_{\mu \nu}$ is given by
\begin{align}
\delta h_{\mu \nu} = \del_{\mu} \xi_{\nu} + \del_{\nu} \xi_{\mu}
\end{align}
where $\xi_{\mu}$ is the gauge parameter.
This is nothing but the diffeomorphism transformation.
This gauge transformation is sometimes called the ``vector type'' in the
literature of fractons (see Appendix~\ref{sec:dual_vector}).
The gauge invariant field strength of $h_{\mu \nu}$ is given by the linearized Riemann
tensor containing two spacetime derivatives \cite{Hull:2000zn}.
On the other hand, the symmetric tensor $A_{\mu \nu}$ that we have discussed
has the gauge symmetry \eqref{eq:scalar_gauge_transformation} called the ``scalar type.''
The gauge invariant field strength is given by \eqref{eq:field_strength}
containing one spacetime derivative.
As we will see in the following, these facts result in the quite
different consequences.

In order to discuss the physical meaning of the mixed symmetric tensor
gauge field that appeared in the dual of the gauge field $A_{\mu \nu}$, we
consider the source term.
It is natural that the type $(2,1)$ field 
couples to a type $(2,1)$ current $J^{\mu \nu | \rho}$.
The Lagrangian is given by 
\begin{align}
\mathcal{L} = 
- 
\frac{1}{6}
\mathcal{F}_{\mu \nu \rho \sigma}
\mathcal{F}^{\mu \nu \rho \sigma}
+ 
\frac{2}{3}
A_{\mu \nu | \rho} J^{\mu \nu | \rho},
\end{align}
where 
$
\mathcal{F}_{\mu \nu \rho \sigma}
$ 
is the field strength of $A_{\mu \nu | \rho}$.
The equation of motion is obtained as
\begin{align}
&
\del_{\sigma}
\Big\{
\mathcal{F}^{\mu \nu \rho \sigma}
+ \mathcal{F}^{\mu \rho \nu \sigma}
+ \mathcal{F}^{\rho \mu \nu \sigma}
+ \mathcal{F}^{\nu \rho \sigma \mu}
+ \mathcal{F}^{\rho \nu \sigma \mu} 
\notag \\
&\qquad 
+ \mathcal{F}^{\rho \sigma \mu \nu}
+ \mathcal{F}^{\nu \sigma \rho \mu}
+ \mathcal{F}^{\sigma \rho \mu \nu}
+ \mathcal{F}^{\sigma \mu \rho \nu}
- (\mu \leftrightarrow \nu)
\Big\}
= \  
-J^{\mu \nu | \rho}.
\end{align}
This immediately results in the current conservation $\del_{\mu} \del_{\nu} \del_{\rho} J^{\mu
\nu| \rho} = 0$ and also the relation 
\begin{align}
\del_{\nu} \del_{\rho} J^{\mu \nu | \rho} = 0.
\label{eq:(2,1)_current_conserve2}
\end{align}
Recalling that a 2-form $B_{\mu \nu}$ naturally couples to a
current $J^{\mu \nu} = - J^{ \nu \mu}$ that
contains the charge density $\rho^{i} = J^{0i}$ for a 1-brane extending along
$x^{i}$, the relation $\del_{j} J^{0i | j}
=\rho^{i}$ implies that $J^{0i | j}$ corresponds to the
density of the dipole moment associated with $\rho^{i}$.
Indeed, assuming $\partial_0\partial_0J^{0i | 0} = \partial_k\partial_0J^{ik | 0} = 0$,\footnote{
Note that this assumption corresponds to $\partial_0\partial_0J^{00} = 0$ in the rank-2 symmetric gauge theory.
}
the current conservation \eqref{eq:(2,1)_current_conserve2} results in the conservation of the charge and the dipole moment for the
1-brane:
\begin{align}
\frac{{\dop}\mathcal{Q}^i}{{\dop}t} = 
\frac{{\dop}}{{\dop}t} \int \rho^{i} \, {\dop}V = 0,
\qquad
\frac{{\dop}\mathcal{Q}^{i | j}}{{\dop}t} =
\frac{{\dop}}{{\dop}t} \int x^{j} \rho^{i} \, {\dop} V = 0.
\end{align}
Here $\mathcal{Q}^i = \int \! \rho^{i} \, {\dop} V$ and $\mathcal{Q}^{i | j} = \int \! x^{j} \rho^{i}
\,{\dop}V$ is the total charge and the dipole moment.
These observations result in the immobility of the 1-brane.

It is straightforward to generalized the discussion to the cases for
$p$-branes.
The $(p+1)$-form $C_{\mu_1 \cdots \mu_{p+1}}$ naturally couples to the
$p$-brane current $J^{\mu_1 \cdots \mu_{p+1}}$.
The $p$-brane extending along the directions $(x^{i_1}, \ldots,
x^{i_p})$ has the charge $J^{0 i_1 \ldots i_p} = \rho^{i_1 \cdots i_p}$.
The mixed symmetric tensor field $A_{\mu_1 \cdots
\mu_{p+1} | \rho}$ of the type $(p+1,1)$ naturally couples to the dipole current $J^{\mu_1
\cdots \mu_{p+1} | \rho}$.
The equation of motion may be given by
\begin{align}
\del_{\sigma} \mathcal{F}^{[\mu_1 \cdots \mu_{p+1}] \mu_{p+2} \sigma} =
-
 J^{\mu_1 \cdots \mu_{p+1} | \mu_{p+2}},
\label{eq:eom_mixed}
\end{align}
where $\mathcal{F}^{[\mu_1 \cdots \mu_{p+1}] \mu_{p+2}}$ is the 
totally antisymmetrized gauge invariant field strength of the gauge field $A_{\mu_1 \cdots
\mu_{p+1} | \rho}$.
The gauge transformation is given by
\begin{align}
\delta A_{\mu_1 \cdots \mu_{p+1} | \rho} 
= \del_{\rho} \del_{[\mu_1} \lambda_{\mu_2 \cdots \mu_{p+1}]},
\end{align}
where $\lambda_{\mu_1 \cdots \mu_p}$ is the $p$-form gauge parameter.
The equation of motion \eqref{eq:eom_mixed} gives the current conservation 
$\del_{\mu_{p+1}} \del_{\mu_{p+2}} J^{\mu_1
\cdots \mu_{p+1} | \mu_{p+2}} = 0 $.
This results in the conservations of the charges:
\begin{align}
\frac{{\dop}\mathcal{Q}^{i_1 \cdots i_p}}{{\dop}t}
=
\frac{{\dop}}{{\dop}t}
\int \! \rho^{i_1 \cdots i_{p}} \, {\dop}V = 0,
\qquad
\frac{{\dop} \mathcal{Q}^{i_1 \cdots i_p | j}}{{\dop}t}
=
\frac{{\dop}}{{\dop}t} \int \! x^{j} \rho^{i_1 \cdots i_p} \, {\dop} V =
 0,
\end{align}
where 
$\del_{j} J^{0 i_1 \cdots i_{p} | j} = \rho^{i_1 \cdots i_p}, $
$\mathcal{Q}^{i_1 \cdots i_p} = \int \! \rho^{i_1 \cdots i_p} \, {\dop}V$ 
and
$\mathcal{Q}^{i_1 \cdots i_p | j} = \int \! x^j \rho^{i_1 \cdots i_p} \, {\dop}V$
are the
charge density, 
the total charges and the dipole moment for the $p$-branes extending along the
$(x^{i_1}, \ldots, x^{i_p})$-directions.

These relations guarantee the immobility of $p$-branes.
The symmetric tensor gauge field $A_{\mu \nu}$ of rank-2 leads to the
conservation of the dipole moment density for 0-branes when viewed in hindsight.
In $D$ spacetime dimensions, we find the duality between immobile
charged objects (Table \ref{tb:dual_fractons}).

\begin{table}[t]
    \centering
    \begin{tabular}{|c|c|c|}
        \hline
        \textbf{dim} & \textbf{dual fields} & \textbf{charged objects} \\
        \hline
        4 & $\tilde{A}_{\mu \nu}$ & fractons \\
        5 & $\tilde{A}_{\mu \nu | \rho}$ & fractonic 1-branes \\
        6 & $\tilde{A}_{\mu \nu \rho | \sigma}$ & fractonic 2-branes \\
        \vdots & \vdots & \vdots \\
        $d$ & $\tilde{A}_{\mu_1 \cdots \mu_{D-3} | \nu}$ & fractonic $(D-4)$-branes \\
        \hline
    \end{tabular}
    \caption{The dual gauge fields and their corresponding charged
 objects that are dual to the fractons in diverse dimensions.}
\label{tb:dual_fractons}
\end{table}

The analysis allows us to discuss the generalization to the multipole case.
The mixed symmetric tensor $A_{\mu_1 \cdots \mu_{p+1}| \nu_1 \cdots \nu_q}$
of the type $(p+1,q)$ is responsible for the $q$-th order multipole for the $p$-brane charge.
As we have discussed in the previous section, 
the type $(D-3, k-1)$ fields appear as the dual
of the rank-$k$ symmetric tensor field $A_{\mu_1 \cdots \mu_k}$ in $D$ dimensions.
Then there are dualities between a fractonic 
$(k-1)$-th order multipole 
0-brane and a
fractonic $(k-1)$-th order multipole $(D-4)$-brane in $D$ dimensions.

\section{Self-duality conditions and BPS equations}
\label{sec:BPS}

In this section, we study the relation between the self-duality and BPS conditions.
From the relation \eqref{eq:dual_relation}, we define the dual tensor
of $F_{\mu \nu \rho}$ as 
\begin{align}
\star {F}_{\mu \nu \rho} = 
\frac{1}{3} 
\Big(
\varepsilon_{\alpha \beta \nu \rho} F_{\mu} {}^{\alpha \beta}
+
\varepsilon_{\alpha \beta \mu \rho} F_{\nu} {}^{\alpha \beta}
\Big),
\end{align}
where we have denoted the dual operation as $\star$.
Note that $\star^2$ is not proportional to the identity as we have discussed.
Once we have a dual tensor, it is now tantalizing to perform the Bogomol'nyi completion.
Let us consider the following action in the Euclidean space:
\begin{align}
S_{\text{E}} = \frac{1}{6} \int \! {\dop}^4 x \,  F_{\mu \nu \rho} F^{\mu \nu \rho}.
\end{align}
By using the following identity 
\begin{align}
\Big(
F_{\mu \nu \rho} \pm \star {F}_{\mu \nu \rho}
\Big)^2
= 
2 F_{\mu \nu \rho} F^{\mu \nu \rho}
\pm
2 F_{\mu \nu \rho} \star \! {F}^{\mu \nu \rho}
-
\frac{1}{2} F^{\alpha} {}_{\alpha \rho} F_{\beta} {}^{\beta \rho},
\end{align}
we have the Bogomol'nyi completion of the form:
\begin{align}
S_{\text{E}} =& \ 
\frac{1}{12}
\int \! {\dop}^4 x \,
\Big(
F_{\mu \nu \rho} \pm \star F_{\mu \nu \rho}
\Big)^2
+
\frac{1}{12 \cdot 2} 
\int \! {\dop}^4 x \, 
(F^{\alpha} {}_{\alpha \rho})^2
\mp 
\frac{1}{6} 
\int \! {\dop}^4 x \, 
F_{\mu \nu \rho} \star \! F^{\mu \nu \rho}
\notag \\
\ge & \ 
\mp 
\frac{1}{6}
\int \! {\dop}^4 x \, 
 F_{\mu \nu \rho} \star \! F^{\mu \nu \rho}.
\end{align}
Interestingly enough, the bound is rewritten as a total derivative of
the Chern-Simons-like term \cite{Bertolini:2022ijb}
\begin{align}
\frac{1}{6} \int \! {\dop}^4 x \, F_{\mu \nu \rho} \star \! F^{\mu \nu \rho}
=
\int \! {\dop}^4 x \,
\eta^{\mu \sigma}
\del_{\nu}
\Big(
\varepsilon^{\alpha \beta \nu \rho} 
A_{\mu \rho} \del_{\alpha} A_{\sigma \beta}
\Big).
\label{eq:Chern-Simons-like}
\end{align}
The BPS equation is given by
\begin{align}
F_{\mu \nu \rho} = \pm \star \! {F}_{\mu \nu \rho},
\qquad
F^{\alpha} {}_{\alpha \mu} = 0.
\label{eq:BPS_eq}
\end{align}
The former is the self-duality condition while the latter is the
traceless condition 
implying $E^i{}_i=0$. Then the fracton becomes the planon. 
Note that the self-duality condition implies the equivalence of the electric and
the magnetic fields:
\begin{align}
E_{ij} = \pm \frac{1}{2} (B_{ij} + B_{ji}).
\end{align}
This is a generalization of the self-duality condition in Maxwell theory.
The self-duality condition leads to the equation of motion:
\begin{align}
\del_{\rho} F^{\mu \nu \rho} = \pm 
\frac{1}{3}
\Big(
\varepsilon_{\alpha \beta \nu \rho}
\del_{\rho} 
F_{\mu} {}^{\alpha \beta}
+
\varepsilon_{\alpha \beta \mu \rho}
\del_{\rho} 
F_{\nu} {}^{\alpha \beta}
\Big)
= 0,
\end{align}
where we have used the Bianchi identity.
On the other hand, the second condition in \eqref{eq:BPS_eq} gives
$\eta^{\mu \nu} \del_{\rho} F^{\mu \nu \rho} = 0$.
In order this to be consistent with the equation of motion, the
traceless conditions should be imposed from the first.
As emphasized repeatedly, the appearance of the extra traceless condition
is traced back to the fact $\star^2 \not= 1$ which is in sharp contrast
with $*_4^2 = 1$ in Maxwell theory.
The traceless condition is also required by the consistency.
For example, by acting $\star$ on both sides in the self-duality
condition in \eqref{eq:BPS_eq}, we have
\begin{align}
\star F_{\mu \nu \rho} = \pm \star^2 \! F_{\mu \nu \rho}
= \pm 
\Big(
F_{\mu \nu \rho} 
+
\frac{1}{6} F^{\alpha} {}_{\alpha \nu} \eta_{\mu \rho}
+
\frac{1}{6} F^{\alpha} {}_{\alpha \mu} \eta_{\nu \rho}
-
\frac{1}{3} F^{\alpha} {}_{\alpha \rho} \eta_{\mu \nu}
\Big).
\label{eq:star_consistency}
\end{align}
Since the right hand side in \eqref{eq:star_consistency} must be $\pm F_{\mu \nu \rho}$, 
the condition $F^{\alpha} {}_{\alpha \mu} = 0$ is necessary.
Then, the self-duality condition is consistent for the planons.
We also stress that the bound of the action is not topological, {\it i.e.} the total
derivative term \eqref{eq:Chern-Simons-like} explicitly depends on the
metric $\eta^{\mu \nu}$.
This is different from Maxwell theory in which no non-trivial
topological solutions to the self-duality condition $F = \pm *_4 \! F$ in
flat space $\mathbb{R}^4$ exist.

A similar BPS bound is found in higher rank gauge theories.
For example, 
assuming the traceless condition $F^{\alpha} {}_{\alpha \mu \nu} = 0$,
we have the Bogomol'nyi bound for the rank-3 action:
\begin{align}
S_{\text{E}} =& \ 
\frac{1}{12}
\int \! {\dop}^4 x \, 
F_{\mu \nu \rho \sigma} F^{\mu \nu \rho \sigma} 
\notag \\
=& \ 
\frac{1}{24}
\int \! {\dop}^4 x \, 
\Big(
F_{\mu \nu \rho \sigma} \pm \star F_{\mu \nu \rho \sigma}
\Big)^2
\mp
\frac{1}{12}
\int \! {\dop}^4 x \, 
F_{\mu \nu \rho \sigma} \star \! F^{\mu \nu \rho \sigma}.
\end{align}
Then we have the BPS equation
\begin{align}
F_{\mu \nu \rho \sigma} = \pm \star \! F_{\mu \nu \rho \sigma},
\end{align}
where the dual is defined by
\begin{align}
\star F_{\mu \nu \rho \sigma}
= \frac{\sqrt{2}}{4}
\left(
\varepsilon^{\alpha \beta} {}_{\rho \sigma} F_{\alpha \mu \nu \beta}
+
\varepsilon^{\alpha \beta} {}_{\nu \sigma} F_{\alpha \mu \rho \beta}
\right).
\end{align}
The action bound is given by 
\begin{align}
S_{\text{E}} \ge \mp \frac{2 \sqrt{2}}{3} \int \! {\dop}^4 x \, 
\eta^{\alpha \gamma}
\eta^{\beta \delta}
\varepsilon^{\mu \nu \rho \sigma}
\del_{\mu}
\left(
A_{\nu \alpha \beta}
\del_{\rho}
A_{\sigma \gamma \delta}
\right).
\end{align}
This is again non-topological.
A similar structure holds even for the rank-$k$ ($k \ge 4$) theories.
\section{Conclusion and discussions} \label{sec:conclusion}
In this paper, we studied the electric-magnetic duality for Lorentz
invariant symmetric gauge theories in diverse spacetime dimensions.
We stress that the structure of the electric-magnetic duality in the symmetric gauge theories
is quite different from that of the ordinary $p$-form electrodynamics.
It is a common knowledge that a $p$-form field theory has the dual description of
the $(D-p-2)$-form theory in $D$-dimensional spacetime.
This is easily understood by the properties of the differential form,
namely, by the Hodge and the external derivative operators.
On the other hand, the duality for the symmetric gauge theories does not
exhibit such kind of structures since there are no 
analogous operations based on the differential calculus.
Indeed, we performed the duality operation by the direct
calculations and showed that the duality for the symmetric gauge
theories arises only in four dimensional spacetime.
This is true even for higher rank symmetric gauge theories for which
immobile multipole objects are implemented due to the generalized 
current conservation law.
In higher dimensions $D > 4$, we showed that the mixed symmetric tensor
fields appear in the dual side of the symmetric gauge
theories. 
The mixed symmetric tensor gauge 
fields are easily understood by the Young tableaux.
We showed that the mixed symmetric gauge field $A_{\mu \nu | \rho}$ of
the type $(2,1)$ Young tableau appears in the dual of the symmetric gauge field $A_{\mu \nu}$
in five dimensions.
The situation is similar to the dual gravitons discussed in the
literature \cite{Hull:2000zn, Hull:2001iu, Casini:2003kf, Hull:2024qpy}.
However, the case we discussed is different from that of the dual graviton in its gauge symmetry.
They naturally couples to the 
mixed symmetric tensor currents that implement the conservation of
the dipole moment associated with the charged extended objects, \textit{i.e.}\ $p$-branes.
We found the gauge invariant actions for the mixed symmetric 
tensor gauge fields. 
The conservation of the dipole moment for $p$-branes follows from the equation
of motion of the mixed symmetric tensor gauge
 fields.
On the other hand, in lower dimensions $D \le 3$, we found there are
no non-trivial electric-magnetic dualities for the symmetric gauge theories.

We also showed that an interesting hierarchical structure of the symmetric gauge
theories. The symmetric rank-2 theory can be seen as a bi-Maxwell
theory and the rank-3 theory looks like a bi-rank-2 theory and
so on. 
The structures are understood by the multi-form calculus discussed in
the literature \cite{Francia:2004lbf, deMedeiros:2002qpr, Hinterbichler:2022agn}.
This hierarchical structure helps us to solve the equation of
motion systematically.
Other issue is physical consequences associated with the mixed symmetric 
tensor gauge fields.
This can be understood by viewing the symmetric gauge field of rank-2 as
a Maxwell field with one extra index.
The current conservation that follows from the equation of motion for
$A_{\mu \nu}$ results in the immobile charged particles, \textit{i.e.}\ $0$-branes.
We showed that the mixed symmetric gauge 
fields that appeared in our dual theories are interpreted as $p$-form fields with one extra index.
Following these facts, 
it is natural that the mixed symmetric tensor of the type $A_{\mu_1
\cdots \mu_{p+1} | \nu}$ describes immobile $p$-branes.
In principle, more extra indices can be added to $(p+1)$-form fields.
The resulting theory is expected to implement the multipole
generalization of the $p$-brane charges.
We will explore this direction more in the 
future works.
Physics associated with the immobility and the fractonic nature of
$p$-branes will be discussed elsewhere.

We also studied the Bogomol'nyi completion of the action for the
symmetric gauge theories in Euclidean space.
We found that the BPS equation is given by the self-duality condition
for the field strength $F_{\mu \nu \rho}$ supplemented by the traceless
condition. This is a generalization of the self-duality condition of
the $p$-form electrodynamics.
We found that the BPS bound 
of the action is given by the total derivative of the
Chern-Simons-like term. Notably, the bound is not topological which is
different from the instantons in ordinary gauge theories.
This fact gives a room for non-trivial solutions to the ``instantons''
in the symmetric gauge theories.
This is in contrast with the instantons in the 
$\group{U}(1)$ Maxwell theory where it admits only the trivial topological structure.

In the formalism presented in this paper, we considered the Abelian gauge
symmetry. It is natural to generalize the formalism to non-Abelian gauge
groups.
Gauge theories based on the fracton gauge principle are relatively new. 
It would be interesting to examine in detail the fundamental field
theoretical properties of this theory 
such as supersymmetrization \cite{Yamaguchi:2021qrx, Katsura:2022xkg, Honda:2022shd}.
We will work on these issues in future studies.

\subsection*{Acknowledgments}
The work of S.~S. and K.~S. is supported by Grant-in-Aid for Scientific
Research, JSPS KAKENHI Grant Number JP25K07324.


\begin{appendix}

\section{Dual of vector type gauge theory}
\label{sec:dual_vector}

In this appendix, we show the duality for the symmetric gauge theory of
vector-type gauge invariance.
The gauge theory presented in the main body is sometimes called the scalar gauge theory since the gauge
parameter $\lambda$ is a Lorentz scalar.
We can consider another symmetry called the vector gauge symmetry.
The vector gauge transformation is given by
\begin{align}
\delta A_{\mu \nu} = \del_{\mu} \lambda_{\nu} + \del_{\nu}
 \lambda_{\mu},
\label{eq:vector_gauge_transformation}
\end{align}
where $\lambda_{\mu}$ is the gauge parameter.
The Lagrangian \eqref{eq:symmetric_maxwell_Lagrangian} is not invariant
under the vector gauge transformation.
The following Lagrangian is invariant under the scalar gauge
transformation \eqref{eq:scalar_gauge_transformation} together with the 
vector gauge transformation \cite{Bertolini:2022ijb}:
\begin{align}
\mathcal{L}_{LG} = - \frac{1}{6} F_{\mu \nu \rho} F^{\mu \nu \rho} +
 \frac{1}{4} F^{\alpha} {}_{\alpha \rho} F^{\beta} {}_{\beta} {}^{\rho}.
\label{eq:LG}
\end{align}

We start from the Lagrangian of the independent fields $F_{\mu \nu
\rho}$ and $\tilde{A}_{\mu \nu}$.
\begin{align}
\mathcal{L} (F,\tilde{A}) =& \ 
-\frac{1}{6} F_{\mu \nu \rho} F^{\mu \nu \rho}
+
\frac{1}{4} F^{\alpha} {}_{\alpha \rho} F^{\beta} {}_{\beta} {}^{\rho}
+
\frac{2}{3}
\tilde{A}^{\alpha \beta}
\varepsilon_{\alpha} {}^{\mu \nu \rho}
\del_{\mu} F^{\beta \nu \rho}.
\end{align}
By integrating out the field $\tilde{A}_{\mu \nu}$, we have the solution \eqref{eq:rank2_Bianchi_solution}
and the theory is given by the Lagrangian \eqref{eq:LG}.
By integrating out $F_{\mu \nu \rho}$, we have
\begin{align}
F_{\mu \nu \rho} 
- \frac{1}{6} \eta_{\mu \nu} F^{\alpha} {}_{\alpha \rho}
= 
\frac{1}{3} 
\Big(
\varepsilon_{\alpha \beta \nu \rho} \tilde{F}_{\mu} {}^{\alpha \beta}
+
\varepsilon_{\alpha \beta \mu \rho} \tilde{F}_{\nu} {}^{\alpha \beta}
\Big),
\label{eq:vector_dual}
\end{align}
where $\tilde{F}_{\mu \nu \rho} = \del_{\mu} \tilde{A}_{\nu \rho} +
\del_{\nu} \tilde{A}_{\mu \rho} - 2 \del_{\rho} \tilde{A}_{\mu \nu}$.
By multiplying with $\eta^{\mu \nu}$ both sides in
\eqref{eq:vector_dual}, we obtain $F^{\alpha} {}_{\alpha \rho} = 0$.
We find that the trace part vanishes and the dual Lagrangian is again given by
\eqref{eq:dual_Lagrangian}.

\end{appendix}


\end{document}